\documentclass[preprint,11pt]{elsarticle}
\usepackage[utf8]{inputenc}
\usepackage{hyperref}       %
\usepackage{url}            %
\usepackage{booktabs}       %
\usepackage{amsfonts}       %
\usepackage{nicefrac}       %
\usepackage{microtype}      %
\usepackage{amsmath}
\usepackage{standalone}
\usepackage{wrapfig}
\usepackage{adjustbox}
\usepackage{siunitx}
\usepackage[table,xcdraw,dvipsnames]{xcolor}
\usepackage{caption}
\usepackage{subcaption}
\usepackage[colorinlistoftodos,prependcaption]{todonotes}
\usepackage{array}
\usepackage{algcompatible}
\usepackage{algorithm}
\usepackage{algorithmicx}
\usepackage[noend]{algpseudocode}
\usepackage{bbm}

\usepackage{tikz}
\usetikzlibrary{positioning}
\usetikzlibrary{arrows}
\usetikzlibrary{shapes.geometric}
\usetikzlibrary{trees}
\usetikzlibrary{calc}
\usepackage{lscape}

\begin{document}
\begin{frontmatter}

\title{\vspace{-6\baselineskip} COVI-AgentSim: an Agent-based Model for Evaluating Methods of Digital Contact Tracing }\vspace{-2\baselineskip}

\vspace{-3\baselineskip}
\author{\small Prateek Gupta*\footnote{Correspondence to pgupta@robots.ox.ac.uk}, Tegan Maharaj*, Martin Weiss*, Nasim Rahaman*, \\
Hannah Alsdurf**, Abhinav Sharma**, Nanor Minoyan**, Soren Harnois Leblanc**, 
Victor Schmidt,  Pierre-Luc St. Charles, Tristan Deleu, 
Andrew Williams, Akshay Patel, Meng Qu, Olexa Bilaniuk, Gaétan Marceau Caron, Pierre Luc Carrier, Satya Ortiz-Gagné,  Marc-Andre Rousseau, David Buckeridge, Joumana Ghosn, Yang Zhang,  Bernhard Schölkopf, Jian Tang, Irina Rish, Christopher Pal, Joanna Merckx,  Eilif B. Muller, Yoshua Bengio
\vspace{-\baselineskip}
}

\begin{abstract}
\small
\textbf{Background:} The rapid global spread of COVID-19 has led to an unprecedented demand for effective methods to mitigate the spread of the disease, and various digital contact tracing (DCT) methods have emerged as a component of the solution. In order to make informed public health choices, %
there is a need for tools which allow evaluation and comparison of DCT methods.%
\\\textbf{Methods:} We introduce an agent-based compartmental simulator we call COVI-AgentSim, integrating detailed consideration of virology, disease progression, social contact networks, and behaviour/mobility patterns, based on parameters derived from empirical research. We verify by comparing to real data that COVI-AgentSim is able to reproduce realistic COVID-19 spread dynamics, and perform a sensitivity analysis to verify that the relative performance of contact tracing methods are consistent across a range of settings. 
We use COVI-AgentSim to perform cost-benefit analyses comparing no DCT to: 1) standard binary contact tracing (BCT) that assigns binary recommendations based on binary test results; and 2) a rule-based method for feature-based contact tracing (FCT) that assigns a graded level of recommendation based on diverse individual features.%
\\\textbf{Findings:}
We find all DCT methods consistently reduce the spread of the disease, and that the advantage of
FCT over BCT is maintained over a wide range of adoption rates.  Feature-based methods of contact tracing avert more disability-adjusted life years (DALYs) per socioeconomic cost (measured by productive hours lost). %
\\\textbf{Interpretation}: %
This research provides a useful testbed to compare and optimize real-world implementations of contact tracing (CT) schemes, a first step in responsible and informed use of CT as an epidemic intervention tool. Our results suggest any DCT method can help save lives, support re-opening of economies, and prevent second-wave outbreaks, and that FCT methods are a promising direction for enriching BCT using self-reported symptoms, yielding earlier warning signals and 
a significantly reduced spread of the virus per socioeconomic cost.
\end{abstract}


\end{frontmatter}

\section{Introduction} \label{sec:intro}

 While vaccine development is underway, countries facing the COVID-19 pandemic are confronted with a limited choice of non-pharmaceutical interventions to control virus propagation. %
 Mathematical modelling and econometric studies show that lockdown has been effective in containing the COVID-19 pandemic in many settings \cite{lockdown1, lockdown2}. However, sustained limitation of human contact is likely not viable with regards to individuals' mental health \cite{rossi}, financial security \cite{Bonaccorsi15530}, children's development \cite{wang2020mitigate}, or national and global economies \cite{mandel}.
 Improving effectiveness of non-pharmaceutical public health strategies is therefore of paramount importance to mitigate the spread of the disease, at least in the short and medium term.

\textbf{ Contact tracing (CT)} is a public health strategy for mitigating the spread of infectious diseases, and consists of alerting individuals that they have recently been in close contact with an infected person (hereafter, an `index case') and taking preventative measures.
In \textbf{manual contact tracing (MCT}), this process is undertaken by public health workers who telephone, email, and/or interview confirmed index cases and subsequently do the same for all persons the index case has been in contact with during their suspected window of infectiousness (hereafter, `contacts').  
 While MCT has been used successfully to contain previous respiratory infection epidemics (e.g. SARS in 2003), the massive scale of the COVID-19 pandemic has overwhelmed the capacity of many public health departments undertaking MCT \cite{watson2020national}. The efficacy of MCT is also limited by peoples' recall of their contacts, which is problematic especially when asymptomatic or presymptomatic transmission plays a large role, as appears to be the case for COVID-19 \cite{he2020temporal, luo}.
 
\textbf{Digital contact tracing (DCT)}, which uses smartphone apps to send messages between contacts with little to no delay, has been successfully deployed in some settings (e.g. Singapore and South Korea \cite{watson2020national}) to supplement the limited manpower available for MCT. 
By enabling large-scale processing of individual-level data (e.g. on testing, presence of symptoms), facilitating communication with potentially-exposed individuals (e.g. through phone alerts), and allowing for non-identifying location-based tracing, DCT has the potential to both maximize efficiency of CT, as well as provide real-time data to inform understanding of COVID-19 and public health decision-making. 
DCT, however, is not free of disadvantages. Due to a wide range of privacy concerns about smartphone communications, DCT suffers from poor adoption by the public \cite{Rimpilainen2020rapid}.

Additionally, most  countries using DCT have adopted a simple form which informs and recommends quarantine to all digitally-recorded contacts of cases confirmed through testing. We call these systems Binary Contact Tracing (BCT) because they recommend users either to quarantine or not (binary decisions) based on whether a past contact took place with a confirmed index case (binary input feature). 
COVID-19 is a challenging disease to mitigate with BCT for two primary reasons (i) BCT currently relies on \textbf{reverse transcriptase PCR} (RT-PCR) tests which have high disease phase-dependent false negative rates. To make it worse, these tests are expensive, and may require a long time to obtain results \cite{kucirka2020variation, guglielmifast} (ii) the  majority of transmissions of SARS-CoV-2 take place before the infector shows any symptoms, thereby reducing the likelihood that a potential infector would have been tested before transmission \cite{cdc2020planning}. 

We observe that there are a wide variety of clues potentially available to a contact tracing app that would allow for non-binary, individualized recommendations, thereby offering significant improvements to BCT.  We call these methods \textbf{feature-based contact tracing (FCT)}, and hypothesize they could provide an important and effective means of reducing the spread of the disease, perhaps even more effectively than BCT at lower adoption rates.

Recognizing this potential, we propose COVI-AgentSim - a software testbed\footnote{Code available here: https://github.com/mila-iqia/COVI-AgentSim} to design, evaluate and benchmark DCT methods using cost-benefit analysis in terms of lives saved, reduction in effective reproductive number ($R_t$) of the virus, disability-adjusted life years (DALYs) averted, and productive hours lost. By using an agent-based model (ABM) as the foundation of this testbed, we are able to simulate a rich set of individual-level input features.
COVI-AgentSim can be adapted to a region of interest by providing appropriate demographics and contact pattern information for that region.
It can then be calibrated to match published data for that region of interest.

We calibrate COVI-AgentSim to reproduce COVID-19 case and hospitalization data for the region of Montreal, Canada. In order to ensure the simulator is a fair and reliable testbed, we also check that the relative ordering of methods is preserved across wide ranges of simulator parameters and over several metrics. We propose a simple rule-based FCT method which leverages individual-level features to make non-binary recommendations, and compare this approach to BCT and compare both to no-DCT via cost-benefit analyses. We find that both BCT and FCT methods are able to reduce spread of the disease, and our results echo those of recent research \cite{lowadoption} suggesting that DCT methods can still save lives even at low adoption rates. We find evidence that FCT approaches, which leverage rich individual-level features to make graded recommendations, are promising for improving DCT even further.  

Additionally, by stratifying DALYs over age groups, we observe the most DALYs averted per person for those over 80 years of age, even with low app adoption rates in that age group, thus showing the protective effects of younger people using DCT. These results are conservative in estimating the benefits for the most vulnerable populations, since we randomly assign DCT app usage proportional to smartphone usage, yet more vulnerable people (or those close to more vulnerable people) may be more likely to use DCT. Our results thus strongly support the usage of DCT methods as a component of effective public health strategies, and we hope COVI-AgentSim will be a useful resource for development, benchmarking, evaluation, and improvement of DCT methods.

\section{Related work} \label{sec:relwork}

\subsection{Agent-based epidemic modeling} Agent-based models (ABMs) are frequently used to study geospatial and other patterns of disease which vary at an individual level (e.g. \cite{perez2009, macal2009}). They are thus often useful for studying differential effects of policy decisions and interventions on different subgroups of  the population; for instance \cite{hoertel2020} use an ABM to study which post-lockdown measures most effectively protect the most vulnerable, in terms of disease incidence, mortality, and ICU occupancy,  and \cite{aleta2020modelling} and \cite{cencetti2020using} study patterns of COVID-19 spread in different representative locations (university, workplace, and highschools), and the impact of different intervention scenarios in each of these locations. One of the interventions studied by these works is DCT, and both find evidence that it can reduce ICU admissions and help curb the spread of the disease. ABMs are also useful for modeling the impact of outlier individuals or events such as super-spreading, e.g. \cite{kim2018agent}, which are not easily captured by mathematical models of population-level spread.

\subsection{Comparing contact tracing methods} 
Several works have studied the use of smartphone apps in epidemic management, e.g.~\cite{hirsch2018grippe, beckman2014ISIS}, and some work has begun specifically on COVID-19. For instance, Ferretti et al \cite{ferretti2020quantifying} propose 
a mathematical model of infectiousness based on early epidemic data in China and compare binary contact tracing to manual tracing. They quantify the contribution of different transmission patterns (infection through symptomatic individuals, presymptomatic individuals, and from the environment) and the requirements for effective contact tracing. Assuming a 3-day delay in notification (and thereby quarantine of the individual), 
the authors demonstrate MCT could not bring $R_t$ below 1 and hence, could not control the epidemic. Instantaneous contact tracing by a digital tracing application on smartphones could do so ($R_t <$1). Shamil et al. ~\cite{shamil2020} follow a similar approach, but with an ABM taking into account realistic contact patterns, studying the potential efficacy of BCT in controlling the spread of the disease. They find strong dependence of the efficacy of digital contact tracing on app adoption, suggesting that BCT alone is insufficient to control a pandemic unless over 60\% of the population is using the app. \cite{lowadoption} find similar results, emphasizing however that even at very low adoption rates DCT is able to save lives. This suggests that DCT should be considered an important component of public health strategy for mitigating COVID-19. We find similarly that BCT and FCT are unlikely to control a pandemic on their own at low adoption rates (see Section \ref{fig:sensitivity-adoption}), showing that even at 60\% adoption rate these methods must be combined with other strategies to contain the disease. 

Perhaps most similar to our work is Hinch et al. \cite{hinch2020effective}, who also propose an open-source ABM which allows manual and digital contact tracing methods to be compared, with benefits stratified across age. Developed concurrently to our simulator, similarities between the two approaches highlight the importance of several design decisions made independently but converging to the same solutions, e.g. the use of ABMs, a python interface, and the need for empirical testbeds of this nature.  A key difference in our simulator is the rich set of individual-level features (including e.g. pre-existing medical conditions), which allow us to benchmark feature-based contact tracing methods, and also allow for stratification over a larger variety of subgroups. The cost of this level of detail is computation; our simulator models smaller populations at higher fidelity for the same computational budget. We perform a scaling analysis (Appendix \ref{sec:scaling}) in order to ensure the dynamics we produce on these smaller populations are representative of larger populations. However, the simulator of \cite{hinch2020effective} may be preferable for studying binary-only contact tracing methods, or when faster computation is needed relative to individual-level detail.

\section{COVI-AgentSim Overview} \label{sec:abm}

The simulator is an agent-based compartmental model \cite{abm-intro} implemented in Python \cite{pythonmanual} and C \cite{kernighan2006c}, using Simpy \cite{simpy}, a process-based discrete-event simulation framework. For each agent the simulator tracks transitions through Susceptible, Exposed, Infectious, and Recovered (SEIR) states, as well as a variety of individual characteristics, including pre-existing medical conditions, self-reported symptoms, and test results. This rich set of individual features enable the simulation of contact tracing apps which make use of such features.
At the same time, we parameterize our simulator using real-world data when available, and when no data is available we make weak assumptions and investigate the sensitivity of our results with respect to these assumptions (see Section \ref{sim_sensitivity}).

\subsection{Demographics}
COVI-AgentSim simulates the spread of the SARS-CoV-2 virus in a city through contagion events between agents. 
Simulator is initialized with a synthetic population along with the mobility and contact patterns informed by census and empirically derived data.
It can be configured easily for any region of interest (see Appendix~\ref{customizing-the-sim}).
Each agent $i$ in the simulation has individual characteristics (e.g. age, sex, pre-existing medical conditions) denoted $\mathbf{X}_{i}$.
Dwelling characteristics, workplace association, and contact patterns are derived from age-stratified surveys and empirical studies (see Appendix~\ref{app:dwelling}).

\subsection{Mobility Patterns and Disease Spread }
At start of a simulation, a fraction $\alpha$ of the agent population is randomly exposed to COVID-19.
Infection spreads through communities via contagion events at households, workplaces, schools and other random locations. 
Agents move around the city transitioning between locations like households, workplaces and other locations. 
The pattern of each individual's mobility (i.e., which locations they visit, how often they visit them, and how they interact with other individuals at these locations) is set according to~\cite{statista2020, statscansocialconnections2020}.
While at a location, agents sample contacts according to age-stratified contact matrices derived from~\cite{Prem2017ProjectingSC}. 
A detailed discussion of agents' mobility patterns and location dependent contact pattern is provided in Appendix~\ref{app:mobility}.
Figure~\ref{fig:all-contacts-main} compares simulated age-stratified contact patterns with surveyed matrices. 

\begin{figure}[!htp]
    \centering
    \includegraphics[width=\textwidth]{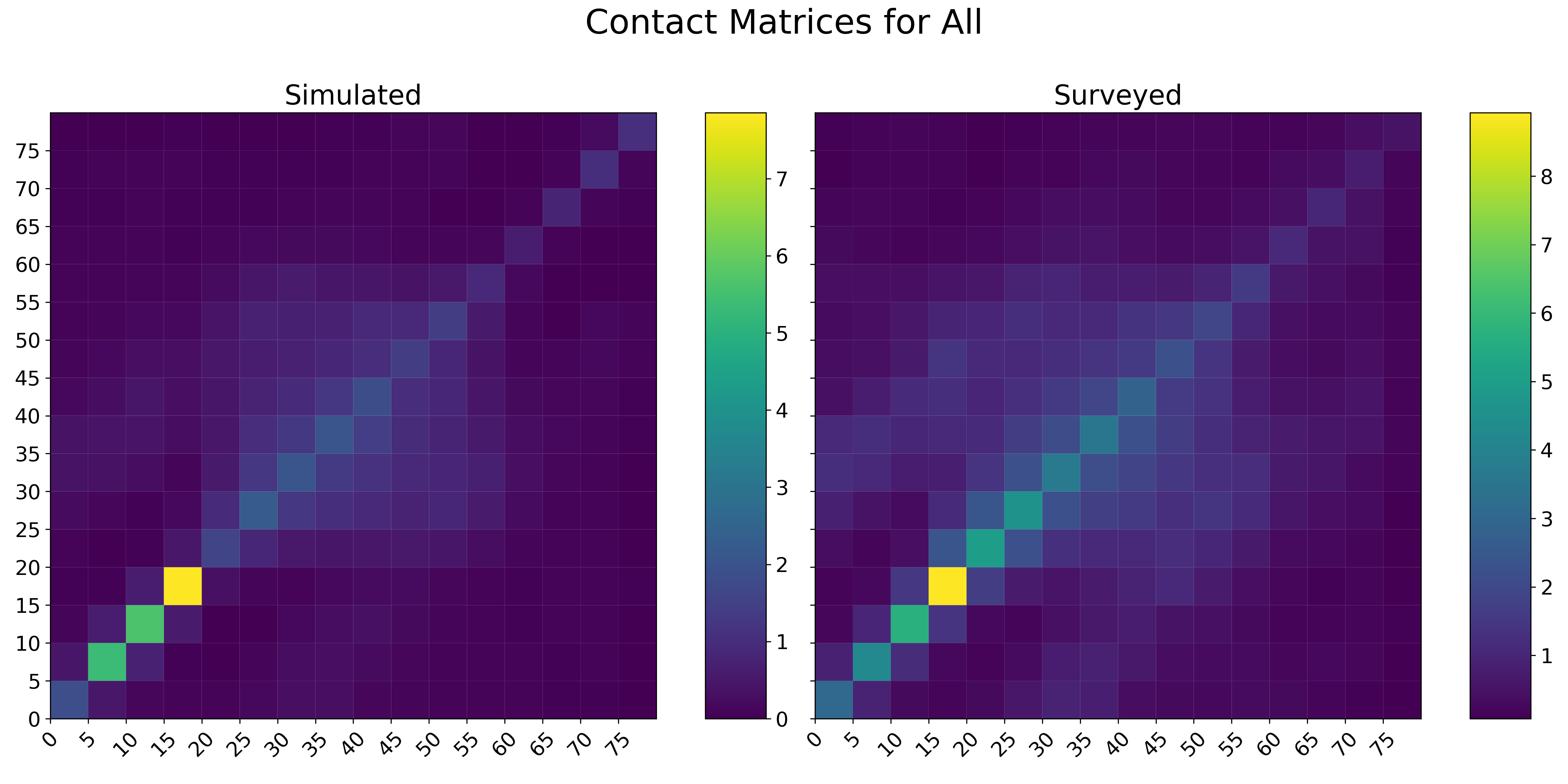}
    \caption{Overall simulated contact pattern (left) yield a similar pattern to empirically derived matrix (right). For location specific contact patterns and mobility of agents see Appendix~\ref{app:mobility}.}
    \label{fig:all-contacts-main}
\end{figure}

\subsection{Transmission Characteristics} Virus transmission takes place anytime an infectious and a susceptible individual are within 2 meters of each other for at least 15 minutes, thereby possibly transmitting viral load to a newly infected agent.  
We model the probability of COVID-19 transmission $P$ according to~\cite{ferretti2020quantifying}.
Borrowing notation from the authors, briefly, this probability is proportional to age-dependent susceptibility $S_a$ of the susceptible agent with age $a$, location-dependent multiplicative factor $B_n$ (for location $n$), symptom status (asymptomatic, mild, severe) dependent ratio $A_s$ of the infectious agent, and a surrogate for cumulative viral load (\textit{EVL}) transmitted from the infectious agent for duration $\delta t$. 
We discuss EVL in the next section. 
A proportionality factor $r$ is used to calibrate the reproductive number of the disease spread.
For the sake of completeness, a mathematical form of this transmission model is presented in Eq~\ref{eq:transmission} and Eq~\ref{eq:prob_transmission}.

\begin{equation}
    \label{eq:transmission}
    \lambda(\delta t, S_a, A_s, n) = \frac{r S_a A_s B_n}{\bar{I}} \int_{\delta t} \textit{EVL},
\end{equation}
\begin{equation}
    \label{eq:prob_transmission}
    P(\delta t, S_a, A_s, n) = 1 - e^{-\lambda(\delta t, S_a, A_s, n)},
\end{equation}
where $P(\delta t, S_a, A_s, n)$ is the probability of the contagion event. 
We use the same values for constants as used by authors in their open-sourced code\footnote{\url{https://github.com/BDI-pathogens/OpenABM-Covid19/blob/master/documentation/parameters/infection_parameters.md}}.

\subsection{Individual Disease Trajectory} 
After such a contagion event, we sample for the infected agent the variables controlling the course of the disease, including symptoms and severity. 
Additionally, we sample a time-series of a quantity proportional to viral load as measured by \cite{goyal2020potency, lauer2020incubation}, which we call \textit{effective viral load} $EVL$; this quantity represents the interaction of the virus with the host's immune system.
We further discuss ($EVL$), its dependence on age and preexisting conditions, and how it is sampled in Appendix~\ref{viral-load}.
A model for sampling symptoms for each agent conditional on whether the agent has cold, flu, or COVID-19 is discussed in details in Appendix~\ref{app:symptoms}.

\subsection{RT-PCR Testing} In COVI-AgentSim, we model RT-PCR testing and its relation to contact tracing applications. As discussed earlier, the disease trajectory for each agent depends on their age and preexisting conditions. 
Thus, there are agents who experience symptoms on a spectrum from none (asymptomatic) to severe: agents who experience more severe symptoms are more likely to seek an RT-PCR test. 
Given the limited testing capacity at the onset of the pandemic, we model a testing facility with a fixed maximum capacity.
For an infectious agent, the outcome of a test is modeled according to disease phase dependent false negative rates as per~\cite{li2020false}.
As an example, 4 days after a SARS-CoV-2 infection, the infected agent will have a false negative rate of 67\%.
Upon receiving a positive test result, the agent is put to self-isolation for $d_{max}$ days with a probability of not following such interaction modeled via dropout parameters. 
We discuss details about testing in Appendix~\ref{app:testing} and hospitalization in Appendix~\ref{app:hospitalization}.

\subsection{Agent Behavior} \label{sec:agent-behavior}
Unlike existing COVID-19 ABMs, our agents are designed to follow varying levels of contact patterns.
For example, number of contacts sampled by an agent in level $k$ at location $l$ corresponds to a fraction ($\gamma^l_k$) reduction in contacts with respect to pre-pandemic number of contacts for that location.
These pre-pandemic number of contacts are available through surveyed studies~\cite{Prem2017ProjectingSC} which we discuss in Appendix~\ref{app:mobility}.
Thus, if there are $n+1$ such levels ranging from $0, 1,..., n$, an agent at location $l$ in level $0$ will draw contacts using pre-pandemic number of contacts ($\gamma^l_k=0$), and an agent at location $l$ in level $n$ is in quarantine i.e. samples no contacts ($\gamma^l_k=1)$.
Fraction $\gamma_k^l$ for intermediate levels $k \in \{1, ..., n-1\}$ are obtained by interpolation scheme (e.g. linear) between $\gamma_0^l$ and $\gamma_n^l$. 
Our choice is motivated by the desire to have a simple model grouping together the effect of choices individuals can make to reduce their likelihood of becoming infected like washing hands, wearing a mask, and physical distancing. 

Each of these levels are further associated with a dropout parameter that represents the fraction of time an agent in level $k$ will drop to level $0$, returning to a higher level of activity (specifically, pre-pandemic numbers of contacts).
COVI-AgentSim can be configured to quarantine individuals due to any of the following triggers (a) confirmed positive test (b) self-reporting of symptoms, (c) recommended by the app, and (d) household member of any of the above cases.
Additionally, in order to model population-wide mobility restrictions we use a Bernoulli distribution with parameter $\beta \in [0,1]$ to sample contacts. 
Thus, an agent that usually draws 12 contacts will now draw $\beta * 12$ contacts on average. 
This modeling choice is a simplification of the varying degrees of government imposed mobility restrictions,
controlled by a population-wide $\beta$.

\subsection{Communication between Agents} \label{sec:agent-communication}
DCT methods rely on capturing high-risk contacts and communicating risk information. 
The population-level app adoption rate is a key parameter because the fraction of contacts captured by this system will be proportional to the square of the app-using fraction of the population. 
To distribute apps throughout our simulated population, we use an age-dependent distribution across smartphone owners. 
We use an age-based breakdown of smartphone users as in~\cite{ferretti2020quantifying}, and use an $UPTAKE$ parameter to vary the population-level adoption rate.
If an agent is assigned an app, there is a provision to report individual characteristics like age and preexisting conditions as well as daily symptoms. 
We further model dropout rates for reporting symptoms as well as a "drop-in" rate for falsely reporting symptoms to account for malicious users and other confounding factors that produce symptom inputs,
such as colds or flu.

When app users $i$ and $j$ have a significant encounter on day $d$ (which can lead to contagion) the CT method can exchange messages between them, on the same day or later.
We have considered two kinds of messages: a contact message exchanged on day $d$ allows $i$ and $j$ to keep track of the encounter event (such as when it happened) as well as a handle allowing them to send each other messages later (potentially via some server), while minimizing breaches in privacy.
A contact message could also contain information about the current degree of contagiousness risk they pose to each other on day $d$.
Later, if $i$ observes new clues suggesting their contagiousness on day $d$ was actually higher, $i$ can send an update message to $j$, with information which can help $j$ update their own evaluation of contagiousness.

We denote $M_{i,j}^d(t)$ the risk message sent on day $t$ by $i$ to $j$ regarding their encounter on day $d$. 
If $t=d$ this is the initial (contact) message, whereas for $t>d$ this is used to update the information $j$ will have at day $t$ about their encounter with $i$ on day $d$. 
Note that $i$ may send multiple update messages to $j$, as they acquire more information about having been infectious on day $d$.
Additionally, we restrict the frequency of communication: if $t'$ is the time of the last risk message sent by $i$ about some encounter, the next message is sent at time $t^{''}$ only if $M_{i,j}^d(t'') \neq M_{i,j}^d(t')$.
The clues which agent $i$ may use to come up with its risk messages include symptoms, test results, pre-existing conditions and received risk messages.
The way to come up with these risk messages, as well as how to adjust behavior based on estimated risk, is specified by the CT method, such as those described below.

\section{Digital Contact Tracing methods}
In digital contact tracing we have two goals: to reduce the individual's spread of disease (by recommending a reduction in mobility to risky individuals), and to inform contacts of additional risk.
Agent's adherence to a recommendation level $k$ entails sampling location specific contact patterns according to that level (see Section~\ref{sec:agent-behavior}). 
We denote agent $i$'s behavior level on day $d$ by $\zeta_d^i$ such that for a total of $n+1$ behavior levels, $\zeta_d^i \in \{0, 1, ..., n\}$.

Determination of this recommendation level is done via a risk estimator that uses a rich suite of features to evaluate agent $i$'s risk history on day $t$. That
risk history is denoted $\mathbf{r}^i_t$, where we constrain risk levels to be non-negative
integers with a maximum value of $R_{max}$. 
We use $r^i_{t, d}$ to denote estimated risk of agent $i$ for day $d$ such that $t - d_{max} < d < t$.
If an agent $i$ had an encounter with an agent $j$ on day $d$ such that $t - d < d_{max}$, $r^i_{t, d}$ is sent to the past contacts as a risk message $M_{i,j}^d(t)=r^i_{t,d}$ if $j$ was a contact
of $i$ on day $d$ (see Section~\ref{sec:agent-communication}).
This determines what information is available to agent's contacts such that they may modify their own behavior or to propagate risk further.

\textbf{Useful Notations}
We use $\mathcal{N}(i)$ to denote a set of agent indices that had at least one digitally recorded contact with agent $i$ in the past $d_{max}$ days.
Further, we use the colon symbol ($:$) to iterate through all possibilities for the variable at that position. 
For example, $M_{i, :}^d(:)$ represents risk messages from agent $i$ to all agents $j \in \mathcal{N}(i)$ for encounter on day $d$, if there was one.
Similarly, $M_{i, :}^:(:)$ represents risk messages sent by agent $i$ to agent $\mathcal{N}(i)$ within the past $d_{max}$ days.

We denote agent $i$'s test result on day $d$ by ${\rm test}^i_{d} \in \{+1, 0, -1\}$, where $+1$ denotes a positive test result, $-1$ denotes a negative test result, and $0$ denotes no test was taken. 
As a simplifying assumption, we assume that a positive test is retained for $d_{max}$ days while a negative test is retained for $d_{min}$ days (we use $d_{min}=2$ in our experiments). 
Thus, an agent $i$ getting a negative test result on day $d$ i.e. ${\rm test}^i_d = -1$ will have this variable set to $-1$ for the next two days i.e. $test^i_{d+1}=-1$, and $test^i_{d+2}=-1$, after which, in the absence of any test $test^i_{d+3} = 0$.
As an input to the {\em Test-based BCT} method, we further denote a history of test results for $d_{max}$ days in the past by a vector $\mathbf{T}^i_d \in \{+1, 0, -1\}^{d_{max}}$. 
We refer to the test result on day $d'$ (where $d - d_{max} < d' \leq d$) as $\mathbf{T}^i_{d, d'}={\rm test}^i_{d'}$ as known on day $d$. 

We similarly denote agent $i$'s symptoms for the past $d_{max}$ days by a matrix $\mathbf{S}^i_d$, such that, with $N_{symptoms}$ categories of symptoms, $\mathbf{S}^i_d \in \{0, 1\}^{N_{symptoms} \times d_{max}}$.
Additionally, agent $i$'s symptoms on day $d - d_{max} < d' \leq d$ are represented via $\mathbf{S}_{d, d'}$.

\subsection{Binary Contact Tracing (BCT)}
\label{sec:BCT}
The most common class of digital tracing methods, Binary Contact Tracing (BCT), can be viewed as a binary classifier with final decision (behavior recommendation) being whether the agent should quarantine or not. 
Most often, the decision boundary is simple: did the individual had a recent contact with someone who received a positive RT-PCR test? 
We refer to these methods as {\em Test-based BCT}, which we describe formally next.

In {\em Test-based BCT} method, for an agent $i$ with ${\rm test}_i^d=+1$, agents $j \in \mathcal{N}(i)$ are notified and recommended to quarantine themselves as discussed in Section~\ref{sec:abm}.
We call this particular method BCT1 because it only affects the immediate contacts (and their household members) of individuals with positive test results; in section \ref{sec:sensitivity} we show results for this baseline, BCT2, where second order contacts may also be quarantined.

\iffalse
\begin{align} \label{eq:bct_risk}
    \mathbf{r}_d, \gamma_d = \begin{cases}
    R_{max} & \forall d > d - d_{max}, 1 &\text{Agent $i$ reports a positive test result} \\ 
    0       & \forall d > d - d_{max}, 1 & \text{$\exists M_{j,i}^d(t') = R_{max}$ where $t' > t - d_{max}$} \\ 
    0       & \forall d > d - d_{max}, 0 &\text{otherwise}
    \end{cases}
\end{align}
\fi

\subsection{Feature-based Contact Tracing (FCT)}
\label{sec:FCT}
We describe a class of methods we call Feature-based Contact Tracing (FCT), which leverage the potentially rich set of features available on a smartphone to make graded recommendations.
As discussed earlier, we make use of following available information on an app to update an agent $i$'s estimated risk history $\mathbf{r}_d^i$ on day $d$: (a) Test results $\mathbf{T}^i_d$, (b) Symptoms $\mathbf{S}^i_d$, (c) Individual characteristics $\mathbf{X}_i$, (d) risk messages $M_{i, :}^:(:)$, and (e) previous estimated risk history $\mathbf{r}_{d-1}^i$.
The agent's estimated risk for the past $d_{max}$ days is then propagated as discussed in Section~\ref{sec:agent-communication}.
In addition to risk estimation, the agent's behavior is set to a level $\zeta^i_d$ based on the estimated risk for that day i.e. $\mathbf{r}_{d, d}$.
In this section, we describe one such rule-based implementation of FCT - \textbf{{\em Heuristic-FCT}} which forms the basis of our experiments. 

\subsubsection{Heuristic-FCT}
With the help of available information about how COVID-19 spreads and manifests itself in an individual in the form of symptoms, we designed a rule-based FCT method. 
Specifically, for every available aforementioned feature type, we determine the agent's risk history for the past $d_{max}$ days.
The agent's risk history on each day $d'$ is taken as the maximum across these per-feature estimated risks.

Broadly, {\em Heuristic-FCT} uses the following rules: 
\begin{itemize}
    \item Test results, $\mathbf{T}_d^i$: the agent's risk is set to $r_{{\rm MAX}} = R_{max}$ if there is a positive test result in the past $d_{max}$ days. 
    This rule takes priority over any other rules
    (assuming that a positive test gives us maximum
    certainty about being in the top risk level).
    
    \item Symptoms, $\mathbf{S}^i_d$:  we identify three categories of symptoms based on how indicative they are of COVID-19. 
    The presence of a highly informative symptom in $\mathbf{S}^i_d$ results in a high risk level $r_{{\rm HIGH}}$; a moderately informative symptom results in a moderate risk level $r_{{\rm MODERATE}}$; and a mildly information symptom results in mild risk level $r_{{\rm MILD}}$. 
    We assign these risk level values for the past $d_{max} / 2$ days in $\mathbf{r}^i_d$, similarly
    to~\cite{hinch2020effective}.

    \item Risk messages, $M_{i, :}^:(:)$: the risk of an agent receiving a risk message $r_{{\rm MAX}}$ is estimated to be $\rho=r_{{\rm HIGH}}$, while one receiving $r_{{\rm HIGH}}$ is estimated to be at $\rho=r_{{\rm MODERATE}}$, and so on. The rationale is that the level of
    risk decreases rapidly as we move away from one agent to 
    its contacts, to the contacts of these contacts, etc.
    We then compute the duration of time when agent $i$ could have been infectious if this contact had caused their infection as $d' + 1 < d'' < d$. 
    Thus, we set the agent's risk to this value $\rho$ until $d''$ days in the past.

    \item Other rules: There are some rules that are designed to override the above rules. 
    For example, an agent with a negative test on day $d'$ is assigned a risk of $0$ from that day onwards.
    Further, an agent is estimated to be at risk level $0$ if there had been no positive test in $d_{max}$ days, no symptoms in the last $d_{max} / 2$ days, and no high, moderate of medium risk messages in a certain past time horizon. 
\end{itemize}

We present a formal description of this risk estimator in Appendix~\ref{app:heuristic-algo}.

\section{Calibration of the Simulator}
\label{sec:calibration}
In this section we seek to provide following evidence: (1) our simulator is producing a reasonable approximation of real-world COVID-19 dynamics, and (2) it is a reliable testbed for comparing contact tracing methods. 
We address (1) by checking the output epidemiological characteristics match published literature (see Figure \ref{fig:SEIR} and Table \ref{tab:epi-char}), and by checking that the hospitalization and mortality statistics  are well-aligned with those found in real world data (Figure \ref{fig:calibration}). We address (2) by performing a sensitivity analysis of the simulator to a wide range of parameters, and checking that the ranking between contact-tracing methods is preserved over these settings, for different metrics.

\subsection{Validation of epidemiological characteristics}

\begin{figure}[htp!]
\centering
\includegraphics[width=.98\columnwidth]{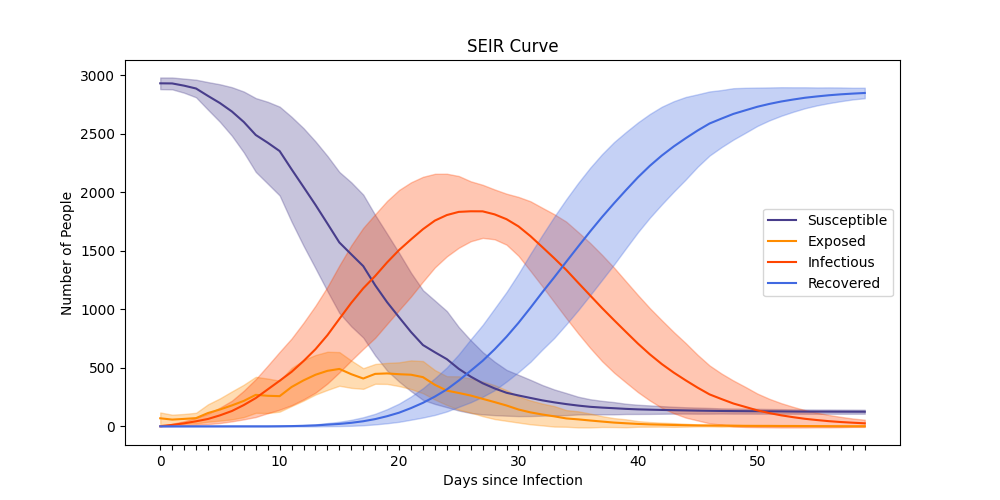}
\caption{
\textbf{SEIR Curves:} (with standard errors) for a population size of 3000 people averaged over 10 random seeds as an output of the proposed simulator.
}
\label{fig:SEIR}
\end{figure}

We calibrate the simulator so that the observed statistics in the simulator are similar to what is observed for COVID-19, plotting SEIR curves in \ref{fig:SEIR} and comparing statistics to published data in \ref{tab:epi-char}.

\sisetup{detect-weight=true,detect-inline-weight=math,detect-mode=true}
\begin{table}[htp!] 
\centering
\begin{tabular}{
    l
    S[table-format=3.3]@{\hspace{5pt}$\pm$}
    S[table-format=2.0]@{\hspace{35pt}}
    l
    }
\hline
\textbf{Metric}                                     &\textbf{Simulator $\mu$ }& \textbf{$\sigma$ }& \textbf{Reference}  \\ \hline
\cellcolor[HTML]{EFEFEF}Incubation                  & 5.505 & 0.01     &  5.807 \cite{lauer2020incubation} \\
\cellcolor[HTML]{EFEFEF}Infectiousness onset        & 3.357 & 0.016    &  2.3  \cite{he2020temporal} \\
\cellcolor[HTML]{EFEFEF}Recovery                    & 18.886 & 0.035   &  14 \cite{he2020temporal} \\
\cellcolor[HTML]{EFEFEF}Generation time             & 4.341 & 0.018    &  3.99 \cite{Knight2020} \\
\cellcolor[HTML]{EFEFEF}Daily contacts              & 16.355 & 0.137 & 13.4 \cite{mossong2008social} \\
\cellcolor[HTML]{EFEFEF}Presymptomatic transmission & 0.563 & 0.011    &  0.44 \cite{he2020temporal} \\
\cellcolor[HTML]{EFEFEF}Asymptomatic transmission   & 0.226 & 0.004    &  0.29 \cite{luo}
\end{tabular}
\caption{\textbf{Simulator statistics:} Key statistics related to Covid-19 epidemiology. Reported numbers are average $\mu$ in days and corresponding 1 standard deviation $\sigma$, computed over 10 random seeds on a population size of 3000. It is important to note that these statistics are a result of the many processes happening within the ABM; there are no parameters that specifically encode these values.}

\label{tab:epi-char}
\end{table}

\subsection{Comparison with real hospitalization and mortality data}
\begin{figure}[htp!]
\centering
\begin{subfigure}{.85\textwidth}
  \centering
  \includegraphics[width=\linewidth]{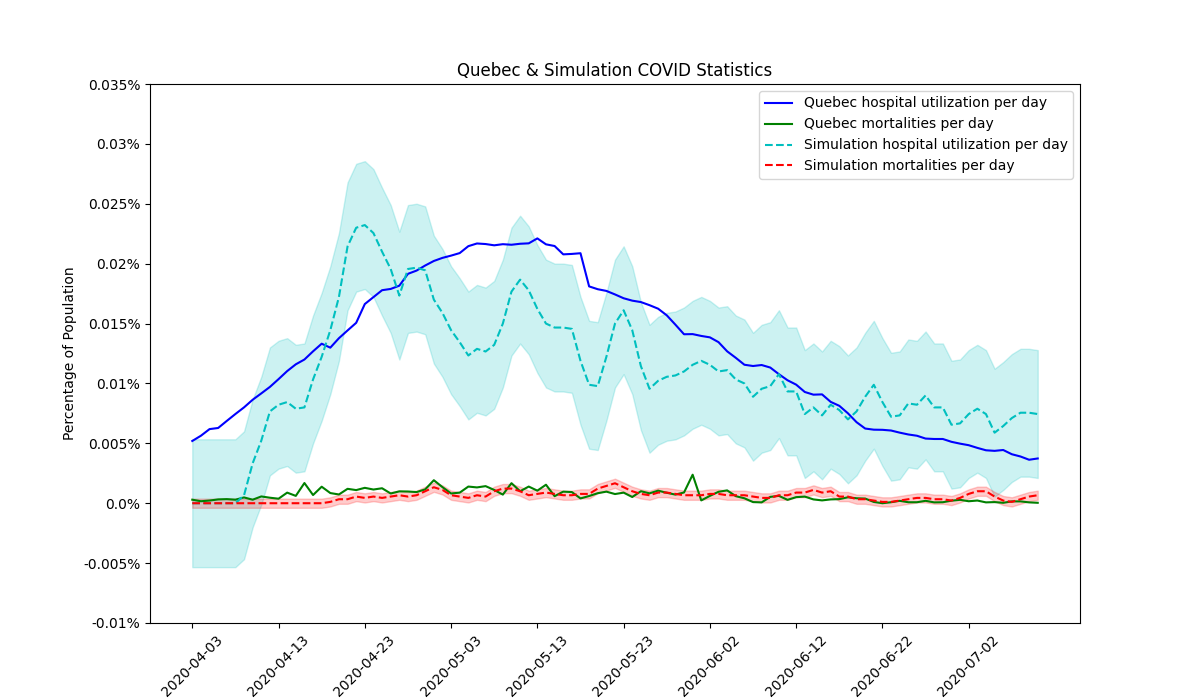}
\end{subfigure}
\caption{\textbf{Comparison of real (solid lines) and simulated (dotted lines) Quebec COVID-19 data.} We compare against the real hospitalization and mortality data (as a percentage of population) during the first wave of COVID-19 in Quebec.  It is important to note that we only report a post-lockdown scenario. We report results with our simulator from 10 runs with a population of 30,000 people. }
\label{fig:calibration}
\end{figure}

We run 10 simulations using random seeds and plot the mean and standard deviation of the hospitalization and mortality statistics over 100 days. We compare these results to the real data reported in Montreal from the same time period. 
Our simulations use 30,000 people and we report results as a proportion of population. We see that under these settings the proportion of the population that is hospitalized or deceased aligns with the data from Montreal (See Figure \ref{fig:calibration}).

\subsection{Sensitivity Analysis} \label{sec:sensitivity}

\label{sim_sensitivity}
A primary contribution of this work is the creation of an ABM which can act as a testbed for comparing COVID-19 contact tracing methods.
While the majority of the parameters in our ABM are chosen according to published literature, much about COVID-19 is still unknown or uncertain. For this reason, we conduct a sensitivity analysis of some key parameters which exhibit high variance across different studies. 

\begin{figure}[htp!]
    \centering
    \includegraphics[width=1.0\columnwidth]{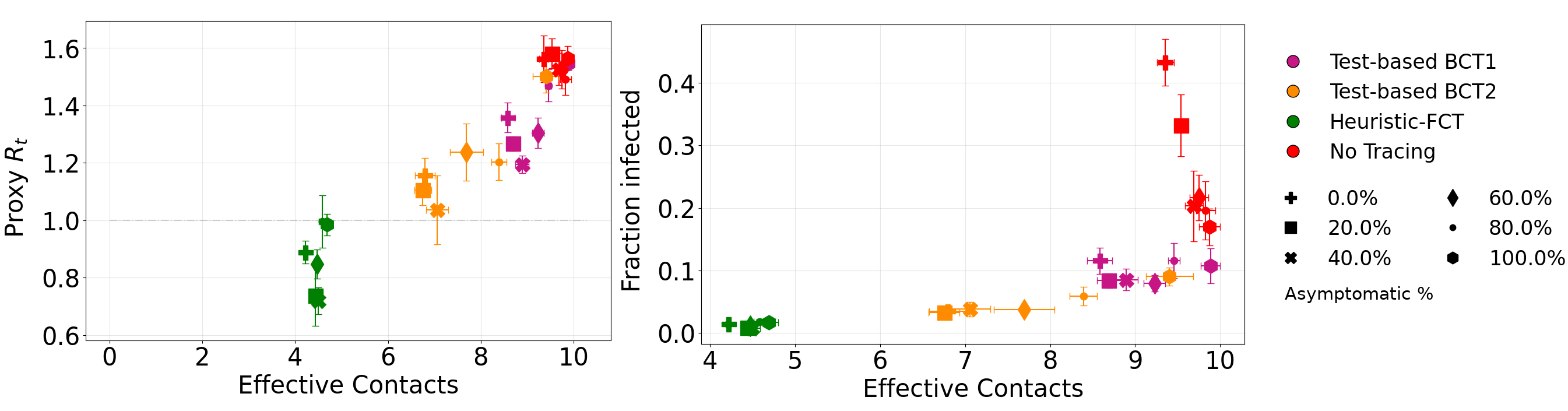}
    \caption{\textbf{Sensitivity to asymptomaticity:} These figures show a comparison between 4 contact tracing methods (including the {\em No Tracing} setting) and varying the proportion of the asymptomatic population between 0 and $100\%$. We plot two views of the same data with different Y-axes: fraction infected and $\hat{R}_t$. The relative performance between the methods are consistent across rates of asymptomaticity and choice of Y-axis. We report the mean and standard deviation across 5 seeds with a population of 3000 people. }
    \label{fig:asymptomatic-rate}
\end{figure}

\begin{figure}[htp!]
    \centering
    \includegraphics[width=1.0\columnwidth]{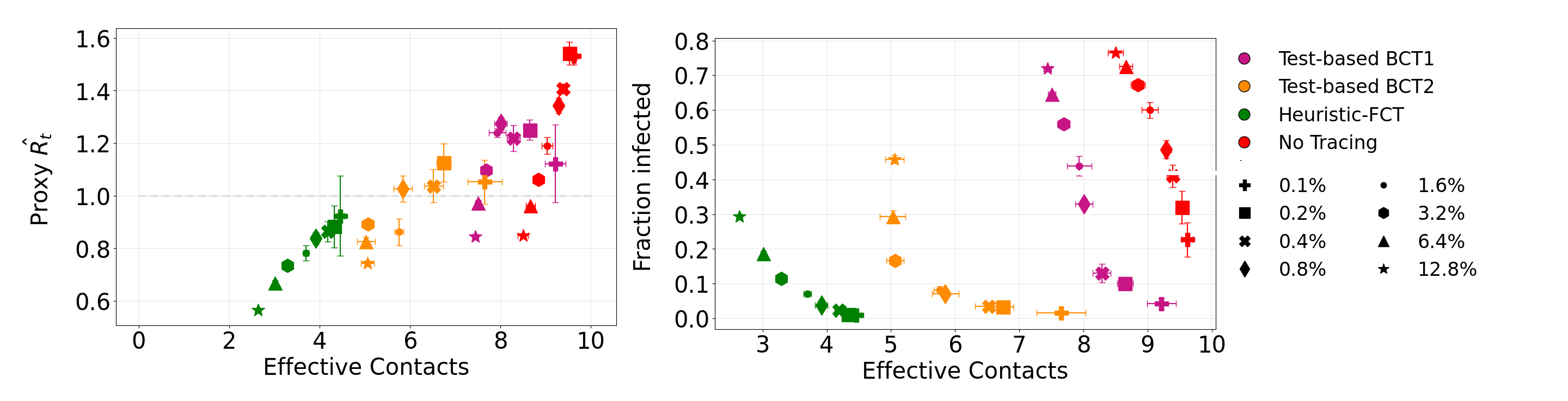}
    \caption{\textbf{Sensitivity to size of initially sick population:} These figures compare 4 contact tracing settings and 8 settings of the proportion of the population initially infected ranging from $.1\%$ up to $12.8\%$. On the left, we see that a larger population of initially sick individuals results in a larger population of infected individuals at the end of the simulation. On the right, we see a counter-intuitive result: the $\hat{R}_t$ at the end of the simulation is \textit{lower} for simulations which started with a larger initially infected population. This can be explained by noting that $\hat{R}_t$ in some sense reports the instantaneous change in rate of spread of the disease while the fraction infected reports the absolute extent of the spread. Crucially, the ranking between the methods remains similar across choice of visualization and initially infected population.  We report the mean and standard deviation for 5 runs under each condition with a population of 3000 people. }
    \label{fig:init-sick-rate}
\end{figure}

Specifically, we study the impact of the asymptomatic proportion of the population (see Figure \ref{fig:asymptomatic-rate}), which is a difficult to measure without widespread serological testing, and the initially infected proportion of the population (Figure \ref{fig:init-sick-rate}).
We observe that the relative efficacy of different methods (i.e. the ranking of methods) holds across a wide range of settings, a desirable characteristic for a comparison testbed.

\section{Experiments} \label{sec:experiments}

\subsection{Setup}
We focus our analysis on the region of Montreal for the period between March and June.
Our agents are initialized with dwelling and workplace characteristics informed from publicly available census data \cite{statcan2016censusmonteral}.
We use $n+1=4$ recommendation levels $\gamma_i$, with $\gamma_0=0$ (full pre-pandemic mobility), $\gamma_4=1$ (full quarantine), $\gamma_3$ as per post-lockdown contact patterns reported in \cite{brisson2020}, and rest of the levels as $\gamma_k = \gamma_{k+1} / 2$.
Given a memory intensive infrastructure required for managing risk messages, we run simulations on a smaller population size of 3000 people.
We initialize each simulation with 0.2\% of the population as infected (6 infections), and run 10 different seeds for each value of $\beta \in \{0.25, 0.275, 0.30, ... , 0.85 \}$ resulting in a total of 420 runs.
These values are chosen so that we can get estimates $\Delta \hat{R}$ of the change in the reproduction number
of the virus, as discussed in Section~\ref{sec:evaluation}.

\subsection{Baselines}
We compare the {\em Heuristic-FCT} method proposed in Section \ref{sec:FCT} with the {\em Test-based BCT} method discussed in Section \ref{sec:BCT}, and a {\em No Tracing} scenario. 
The scenario of {\em No Tracing} corresponds to agents initialized at level $1$ (post-lockdown contacts with some restriction advised) instead of level $0$ (pre-pandemic contacts), in order to compare our methods in the context of the scenario where economies are gradually reopening.
We use {\em Test-based BCT1} and {\em Test-based BCT2} to distinguish between the two variants when necessary, otherwise we use {\em Test-based BCT} to imply {\em Test-based BCT1} which is the method being considered by most of the countries. 

\subsection{Evaluation} \label{sec:evaluation}
We compute the following metrics to evaluate various scenarios:
\begin{itemize}
    \item Average number of contacts per day per human ($C$): We empirically compute the average number of daily contacts per agent in simulation.
    \item Proxy R ($\hat{R}$): We use an empirical calculation to estimate the reproductive number $R$, we call this estimate $\hat{R}$. At any timepoint in the simulation we may approximate $R_t$ by $\hat{R_t}$, by computing the infection tree and taking the ratio of $\frac{\text{number of children}}{\text{number of parents}}$, where parents are recovered agents. This ratio gives an approximate rate of growth of the tree. We use $\hat{R_t}$ at the end of the simulation as our $\hat{R}$.
    \item Daily cases (\%): Percentage of population exposed, i.e., new cases on any given day of a simulation.
    \item Cumulative cases (\%): Percentage of agents in exposed, infectious, or recovered state up until a particular day of a simulation.
    \item Prevalence (\%): Percentage of population in infectious or exposed state on any given day of a simulation
    \item Incidence (\%): Number of agents exposed per 1000 susceptible agents on any given day of a simulation. It is also referred to as attack rate.
\end{itemize}

As the contact tracing methods proposed in this paper change agents' behaviour to varying degrees, it is crucial to compare different methods across similar social mobility restrictions. 
For example, BCT can only use levels $1$ and $n$ while FCT can use all levels from $1$ to $n$.
Therefore, at the same value of $\beta$, BCT will likely under perform as it can not reduce infections by using intermediate levels.
Thus, for a fair comparison of BCT with FCT, we run simulations with varying values of $\beta \in (0, 1)$
and match them for average number of contacts. 

To compare the performance of different methods for the same mobility restriction we empirically compute pairwise difference in mean $\hat{R}$ for a fixed number of contacts $C$. 
This is achieved by obtaining the performance in terms of $\hat{R}$ of a method across a spectrum of values of $C$ (by varying $\beta$), and fitting a Gaussian Process (GP) regression \cite{gpregression} to obtain a functional dependence for each method
between $C$ and $\hat{R}$.
We denote this fitted regression for method $A$ by $\hat{R}^{GP}_A$.
To compute the advantage of method $A$ over $B$, we find $C=x$ for which $A$ yields $\hat{R}^{GP}_A=1$.
Next we compute the {\em advantage} $\Delta \hat{R}_{AB}$ of $A$ over $B$ as $\hat{R}^{GP}_B(x) - 1$.
\begin{equation}
    \Delta \hat{R}_{AB} = \hat{R}^{GP}_B(x) - 1, \text{where,  $x$ is such that } \hat{R}^{GP}_A(x) = 1.
\end{equation}
Note that $R=1$ is the threshold between exponential growth and exponential decay of the virus and serves
as a good point of interest when comparing methods since the objective is to choose a method which brings $R$ below 1,
all else being equal (e.g., general restrictions on social mobility).

\subsection{Comparison of methods} \label{sec:comparison}
\begin{figure}[htp!]
    \centering
    \includegraphics[width=\textwidth]{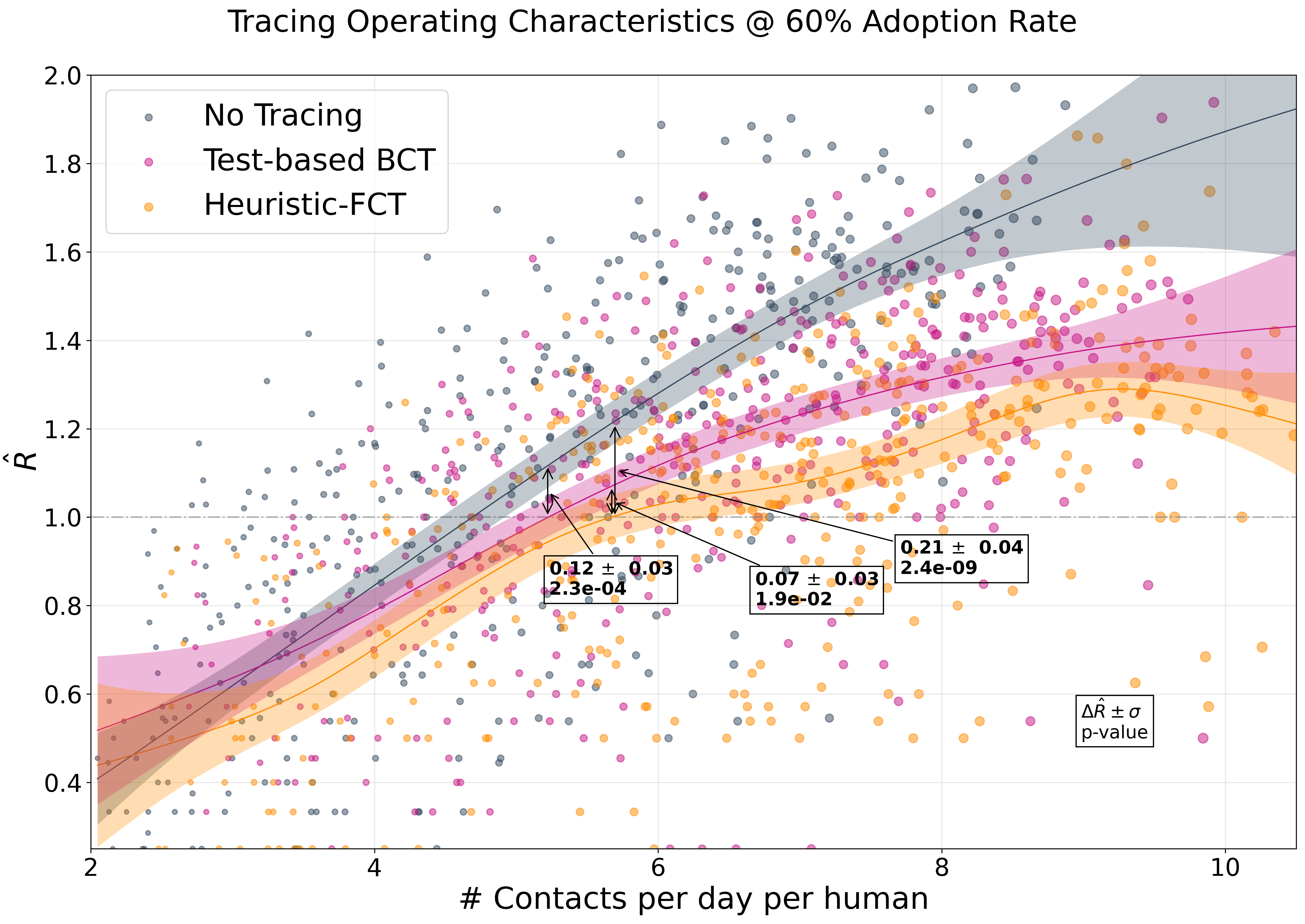}
    \caption{
    \textbf{Pareto front.} We compare DCT methods at 60\% adoption rate, and a No Tracing scenario. For each method, a GP regression is fit as discussed in Section~\ref{sec:evaluation}, approximating a trade-off between mobility and spread of disease. We plot the mean fitted function along with 95\% confidence intervals. Relative to the No Tracing scenario, we observe a statistically
    significant 17.4\% reduction in $\hat{R}$ by Heuristic-FCT as compared to 10.7\% by Test-based BCT method.
    }
    \label{fig:scatter}
\end{figure}

Figure~\ref{fig:scatter} shows a comparison of the GP regression fits between aforementioned DCT methods at 60\% adoption rate, and a {\em No Tracing} scenario, across a range of values of $C$. 
We observe that {\em Heuristic-FCT} significantly improves over {\em Test-based BCT} by reducing $\hat{R}$ by 6.7\%.
Of note is the region around $C = 5.61 \pm 0.5$ where $\hat{R}=1.2$ for {\em No Tracing}. 
We compare the performance of DCT methods in this region to set our comparisons in the context of current scenarios (partial lockdown) where government-imposed restrictions keep $R$ under control.

\begin{figure}[htp!]
    \centering
    \includegraphics[width=\textwidth]{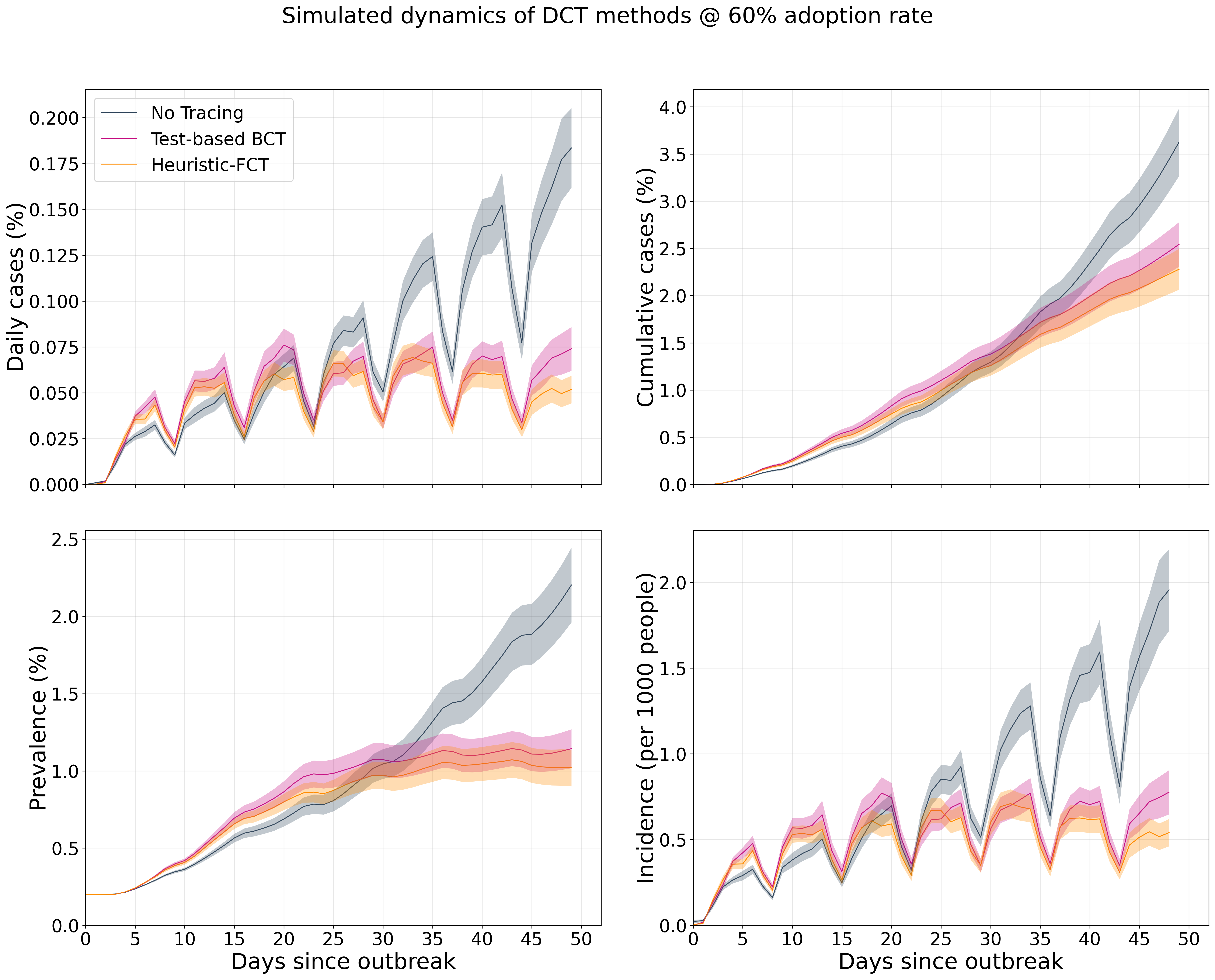}
    \caption{\textbf{Case curves.} We investigate the dynamics of the simulations chosen as per the criterion mentioned in Section~\ref{sec:comparison} to compare the performance of DCT methods at 60\% adoption rate in the context of partial-lockdown. Shown are the mean with one standard error band of the metrics described in Section~\ref{sec:evaluation}. Simulations under Heuristic-FCT exhibit lower attack rates (incidence) as compared to Test-based BCT method, thereby explaining the advantage visible in Figure~\ref{fig:scatter}.}
    \label{fig:case-curves}
\end{figure}

To further investigate the reason for the improvement of {\em Heuristic-FCT} over {\em Test-based BCT}, we peek into the simulations in the concerned region of $C$.
Figure~\ref{fig:case-curves} shows that on average, {\em Heuristic-FCT} exhibits lower incidence as well as prevalence on any given simulation day.
This is expected as {\em Heuristic-FCT} makes use of far richer features as compared to binary test results used by {\em Test-based BCT} to evaluate individual's risk of infecting others.

\begin{figure}[htp!]
    \centering
    \includegraphics[width=\columnwidth]{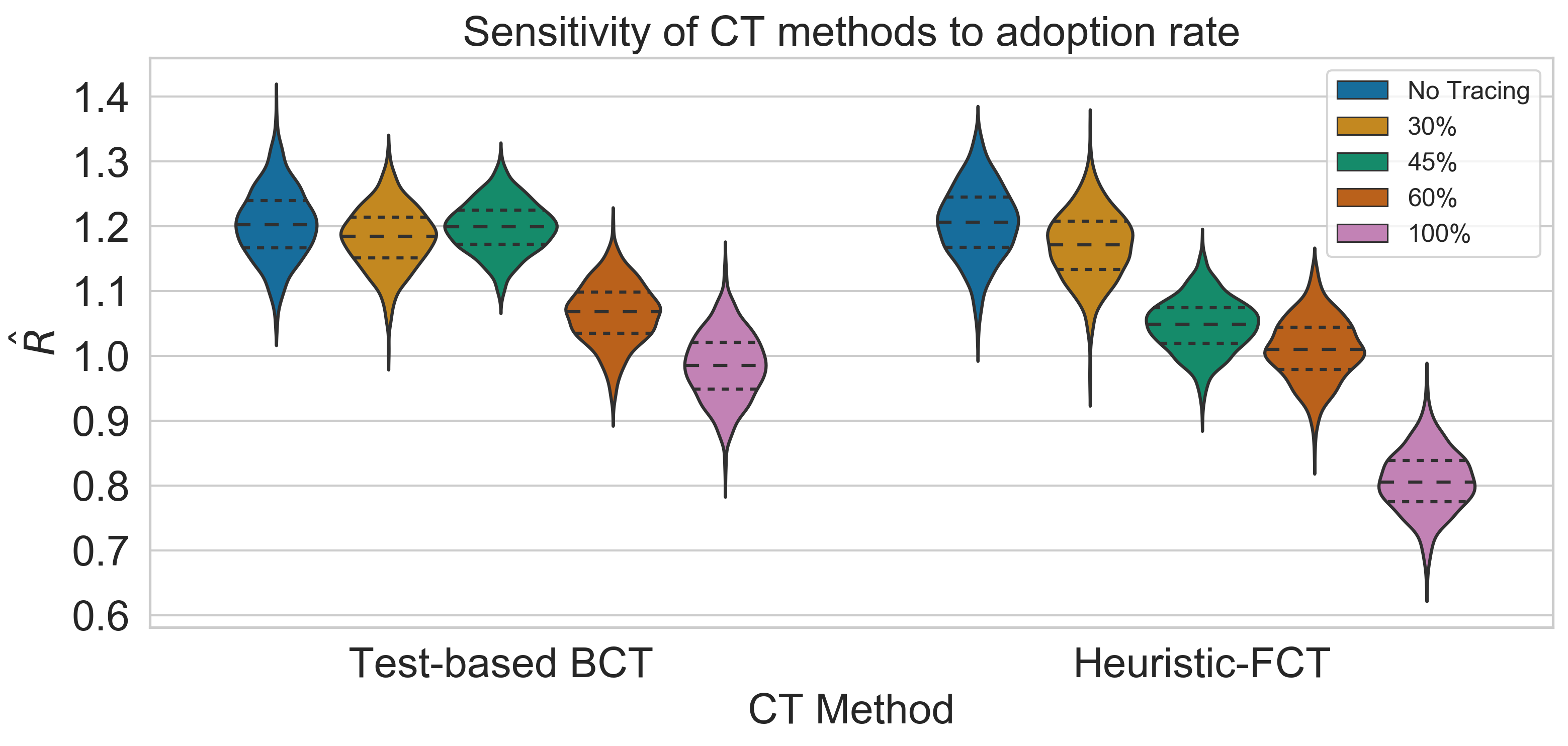}
    \caption{\textbf{Effect of adoption rates.} Performance of DCT methods is heavily dependent on adoption rates. However, Heuristic-FCT retains its advantage over Test-based BCT across this spectrum.}
    \label{fig:sensitivity-adoption}
\end{figure}

Finally, due to network effects, it is likely that adoption rate plays an important role in the performance of DCT methods.
Thus, we evaluate DCT methods at various adoption rates. 
To do this, we run simulations of DCT methods at different adoption rates in a similar way as explained in Section~\ref{sec:evaluation}, and concern our analysis on simulations chosen according to the selection criterion described above.
Figure~\ref{fig:sensitivity-adoption} shows the performance of the considered DCT methods at different adoption rates. 
As expected, the performance diminishes with lower adoption rates, for all the methods. However, we observe that {\em Heuristic-FCT} retains its advantage over BCT methods even at the lower adoption rates.
At the same time, we also note that poor adoption brings both the DCT methods close to {\em No Tracing}, thereby making it more difficult to measure significant advantages over the partial lockdown scenario. 
Thus, we argue that by continuity while at any adoption rate DCT methods can save lives,
adoption rate is crucial for their efficacy.

\section{Cost-benefit analysis}

COVID-19 has been a challenge for health agencies and economic policy decision makers alike. 
Countries around the world have taken unprecedented measures to prevent the collapse of health and economic welfare of people.
Yet, at times, these two objectives have seemed to be at odds with each other in efforts to contain the pandemic.
Therefore, we acknowledge that the evaluation of DCT methods should also stand on sound assumptions of utility maximization in economic theory. In this section, we attempt to relate simulation dynamics to economic outcomes that policy makers can use for decision making. 

To assess the socio-economic burden of COVID-19, we examine the following metrics: (a) Disability-Adjusted Life Years (DALYs) averted \cite{gbd2017}, a measure of lost years of healthy life, and (b) Temporary Productivity Loss (TPL), a measure of economic cost to society of restrictive measures.
Of particular interest is the trade-off between these two measures. 
We use TPL per DALY averted to compare the cost-effectiveness of DCT methods with respect to the {\em No Tracing} baseline. 
This can be thought of as an Incremental Cost-Effectiveness Ratio (ICER), which answers the following question: how much \textit{more} does each unit of additional benefit (in averted DALYs) cost with respect to the {\em No Tracing} baseline?
A breakdown of the methodology used to calculate DALYs and TPL can be found in Appendix~\ref{app:cost-effectiveness}.
One assumption made in computing TPL is that total foregone work hours due to quarantine are scaled by a factor of 0.49\cite{gallacher2020remote} to account for the proportion of agents able to work from home. 

\subsection{Setup}
To assess the impact of the aforementioned CT methods on DALYs and TPL, we run 10 simulations of 3000 agents for each method. 
For a reliable economic analysis, it is necessary to have data on the full trajectory of simulations reaching a post-epidemic steady state comprised of agents that are either susceptible or recovered.
Such a trajectory would help assess whether the DCT methods actually avert the loss of life years, or simply delay them. 
Since CT methods are used to stop individuals from spreading the disease, it is important to know whether the CT methods have successfully reduced the overall proportion of individuals, or if they have simply delayed these infections.

If the simulations are only run for 60 days with 0.2\% infections seeded at the start, then the epidemic has still not reached a post-pandemic steady state, resulting in a possible overestimation of the benefits of the aforementioned CT methods.
Thus, we run longer simulations of 90 days with 5\% infections seeded at the start of simulations.
This particular choice lets us draw preliminary insights into the cost-effectiveness of CT methods without having to scale up the simulations which are extremely compute intensive.

\subsection{Results}
As a first step towards understanding simulation dynamics in terms of healthcare and economic costs, we compute DALYs averted and TPL for each scenario considered in Section~\ref{sec:experiments} under the conditions described above.
We use {\em No Tracing} as a reference scenario to independently evaluate differential benefit of DCT methods over doing nothing.

 \textbf{ICER for No Tracing}Under the assumptions discussed at the onset of this section, our experiments (results in table~\ref{tab:cost-benefit}) suggest that {\em Heuristic-FCT} saves almost six times more DALYs than {\em Test-based BCT}, a significant number of life years saved.
However, we note that the TPL of implementing {\em Test-based BCT} is ~30\% that of {\em Heuristic-FCT}: higher proportions of agents are quarantined under {\em Heuristic-FCT}.
Finally, we note that the cost per DALY averted of{\em Heuristic-FCT} is 58\% of that of {\em Test-based BCT}.
Thus, the cost per healthy year of life saved by {\em Heuristic-FCT} is lower than the {\em Test-based BCT} method. 

\textbf{ICER for {\em Test-based BCT}}To quantify the cost-effectiveness advantage that {\em Heuristic-FCT} provides over {\em Test-based BCT}, we also calculate the ICER of {\em Heuristic-FCT} over {\em Test-based BCT}. 
On average, {\em Heuristic-FCT} saves $(58.12-10.01) = 48.11$ more DALYs than {\em Test-based BCT}. 
However, it costs $(\$755\text{K} - \$223\text{K}) = \$532\text{K}$ more as well.
Therefore, the ICER of {\em Heuristic-FCT} over {\em Test-based BCT} is $\frac{532\text{K}}{48.11} = \$11\text{K}$ per DALY averted.

\begin{table}[htp!]
\begin{tabular}{lcrlrl}
                                & \multicolumn{1}{l}{\em No Tracing}                                & \multicolumn{2}{c}{\em Test-based BCT}                                        & \multicolumn{2}{c}{\em Heuristic-FCT}     \\
\hline
&                                                                            &   \multicolumn{1}{r}{$\mu${\hspace{5pt}$\pm$}}   &  \multicolumn{1}{l}{(SE)}     &  \multicolumn{1}{r}{ $\mu${\hspace{5pt}$\pm$}} & \multicolumn{1}{l}{(SE)}   \\ 

DALYs (years)                               & REF                            & 10.01{\hspace{5pt}$\pm$}                         &  6.96                         & \textbf{58.12}{\hspace{5pt}$\pm$}              &  6.01  \\
TPL (\$)                                    & REF                            & \textbf{\$223K}{\hspace{5pt}$\pm$}               &  1720                         & \$755K{\hspace{5pt}$\pm$}                      &  1868  \\
ICER $\left(\frac{\text{TPL}}{\text{DALY}}\right)$       & REF                            & \$22.3K{\hspace{5pt}$\pm$}                       &  15.5K                        & \textbf{\$13.0K}{\hspace{5pt}$\pm$}            &  1345
\end{tabular}
\caption{Comparison of standard {\em Test-based BCT} and {\em Heuristic-FCT} with {\em No Tracing} as a baseline. Although the TPL incurred by {\em Heuristic-FCT} is higher than that of {\em Test-based BCT}, the TPL per DALY Averted of {\em Heuristic-FCT} is nearly half that of {\em Test-based BCT'}s, meaning a better cost-effectiveness of {\em Heuristic-FCT} than {\em Test-based BCT}. }
\label{tab:cost-benefit}

\end{table}

\begin{figure}[htp!]
\centering

\includegraphics[width=1.\columnwidth]{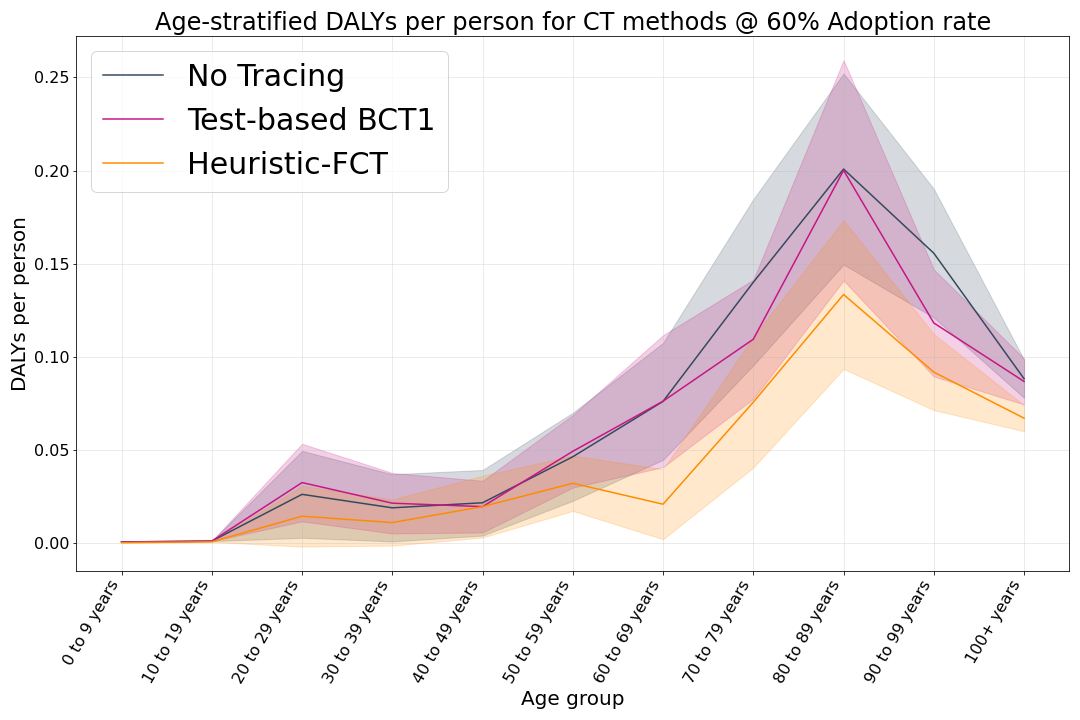}
\caption{
 Impact of different CT methods on DALYs, binned by age group. 
 DALYs per person are total DALYs per age group over the size of each group.
 {\em Heuristic-FCT} consistently outperforms {\em Test-based BCT} across older age groups, most vulnerable to COVID-19.
}
\label{fig:ageDALYs}
\end{figure}

\textbf{DALYs per age group} Since there is evidence in the literature that COVID-19 disproportionately affects the elderly\cite{nurchis2020impact}, we stratify the DALYs per person by age in Figure~\ref{fig:ageDALYs}. It is highest among people aged 80-89 years for both CT methods, which is consistent with the literature. It is worth noting that the mean of {\em Heuristic-FCT} performs at least as well as {\em Test-based BCT} across all age ranges, and notably better for agents aged 60 and over.

\newpage

\section{Limitations}
We attempted to create an easily configurable platform to assist in evaluation of DCT methods and variants thereof. 
Further, we proposed an FCT method that has a potential to improve upon BCT methods, that is widely used in practice. 
However, we would like to point out some of the limitations our work. 

First, although most of the parameters in COVI-AgentSim are informed from published literature, there are assumptions we had to take in the absence of available data. 
In this paper, we have enlisted most of these parameters and corresponding references that informs them. 
It is important to note that as more about COVID-19 will be known these parameters are subjected to change, thereby affecting results in Section~\ref{sec:experiments}.
Additionally, we introduced intermediate levels of user behavior to contrast FCT with BCT.
This is done with the help of introducing a factor $\gamma_n$ for level $n$ that represents fraction reduction in number of contacts relative to pre-pandemic number of contacts ($\gamma_0$), empirical estimates of which have been widely used in epidemiological modeling. 
To obtain intermediate levels, we used interpolation such that $\gamma_k = \gamma_{k+1} / 2$. 
Although it is trivial to experiment with various interpolation schemes, in the absence of user-behavior research, there is not enough that can be done in this regard other than providing the sensitivity analysis of the parameters involved therein.  
In addition to this, we foresee the use of such assumptions to be translated into government policies such that resulting number of contacts can approximate these assumptions.

Second, our modeling framework of FCT relies on certain assumptions of technology which might not hold in practice. 
For example, our assumptions of proximity detection using bluetooth signals~(Appendix \ref{app:bluetooth}) might be unrealistic. 
However,~\citep{briers2020risk}'s work on improving the reliability of bluetooth signal can be a way to address this. 
Further, we assume the app to be active (in foreground) to be able to communicate with nearby phones all the time, an assumption that might not hold depending on the app design.
There are a variety of such technological and ethical considerations which are discussed in~\citep{alsdurf2020covi} to design a successful peer-to-peer FCT app.
COVI-AgentSim incorporates most of the technological assumptions in~\citep{alsdurf2020covi}, however, with additional effort, other assumptions can also be incorporated in COVI-AgentSim to evaluate DCT methods in different settings.

Third, we designed the {\em Heuristic-FCT}
method using rules that were informed by domain knowledge about COVID-19's spread characteristics. 
At this point, we acknowledge that there should be room for improvement in these rules which can be brought upon by amalgamation of ideas in disciplines such as epidemiology, user behavior research, computer science, and statistics to name a few. Alternatively, machine learning methods could be used to learn such rules.
Thus, we think that a unified mathematical framework to analyze FCT methods might help further in development of better FCT methods.

\section{Conclusion} \label{sec:conclusion}

This work presents COVI-AgentSim, a simulation testbed for evaluation of DCT methods.
COVI-AgentSim is an agent-based compartmental model that is initialized with a synthetic population sampled from the census data.
Daily activities as well as interactions for each agent are sampled according to the empirically derived contact matrices.
We calibrate COVI-AgentSim to approximate the spread of COVID-19 virus to the region of Montreal, however, the simulator can be easily configured for other regions via change of an appropriate configuration file.

Finally, we propose the FCT class of contact tracing methods that utilize a richer set of input features as compared to BCT methods (which rely on binary signals like presence or absence of a positive test result). 
In doing this, we aim to provide infected individuals with a warning signal earlier than BCT methods. 
To put FCT in practice, we designed {\em Heuristic-FCT} which uses hand-designed rules to inform an individual's risk 
of infection and infectiousness to others.

Our empirical results show that {\em Heuristic-FCT} results in 6.5\% improvement in $R_t$ 
over BCT methods, and both the methods themselves provide a significant improvement over a partial lockdown scenario.
Experiments with varying adoption rates suggest that the efficacy of DCT methods is heavily dependent upon adoption rates.
It is, however, observed that {\em Heuristic-FCT} retains its benefit over {\em Test-based BCT} method across the adoption rate spectrum, but this advantage was not statistically significant in the face of very low adoption rates, at the scale
of our simulations.

Using an agent-based compartmental model as the foundation of this testbed allows us to simulate a rich set of individual-level features, which we show can potentially be leveraged by DCT methods to improve over the existing BCT methods. 
We hope that the baselines established in this work will encourage and enable the informed development of DCT methods as a first step in their responsible deployment as an epidemic intervention tool, potentially saving lives at lower economic cost during deconfinement and/or second-wave prevention.

Finally, this work joins a growing body of work in considering novel methodologies for rigorous evaluation of interdisciplinary technologies. Epidemiology is fundamentally an intersectional science, touching sociology, biology, behavioural psychology, geography, political science, ecology, mathematical and computational modeling, and many other fields, in a society which is increasingly digitized and globalized. By working together across fields, with careful empirical study, we have hope in dealing with the important issues we face.

\section{Future Work} 
A major direction for future work is to benchmark a wider variety of CT methods, including probabilistic and machine-learning based methods which could make even better use of the features our simulator provides than the {\em Heuristic-FCT} proposed here.
The data generated by our simulator are potentially of interest for training such models to estimate individual-level characteristics which are  predictive of the spread of the disease. 

Our cost-benefit analysis using DALYs and TPL is a first step towards an integrated framework to help policy makers in their decision process.
We see imbuing economically sound decisions in our simulated agents as a step in creating an integrated framework for richer evaluation of DCT methods.

Although COVI-AgentSim is designed to evaluate DCT methods, we foresee directions for the simulator to investigate the impact of a gamut of non-pharmaceutical interventions on containing COVID-19. 
For example, with some work and expertise in manual CT methods, one can compare and evaluate various variants thereof.
Another example is to analyse various COVID-19 testing strategies in conjunction with DCT methods.
Studies of this nature could potentially help public health agencies as well as policy makers in their decision process. 
A far cry, though worth mentioning, is the fact that COVI-AgentSim has essential components for simulating an outbreak, thereby enabling adaptation to other infectious diseases like influenza or tuberculosis.

\newpage
\bibliographystyle{unsrt}
\bibliography{references}

\newpage
\appendix 
\section*{Appendix}
\renewcommand{\thesection}{\Alph{section}}

\section{ABM Demographics}
\label{individual-properties}
Information on the Montreal population regarding age, sex and occupation distribution was retrieved from Canadian Census data~\cite{statcan2016censusmonteral}. 
Prevalences of selected medical conditions considered as COVID-19 infection and prognosis risk factors were determined based on prevalence estimates from nationally representative surveys or medical surveillance programs in Canada: heart disease~\cite{heart_disease, Petrilli2020, williamson2020, killerby2020}, stroke \cite{williamson2020}, asthma \cite{Petrilli2020, williamson2020}, chronic obstructive pulmonary disease (COPD) \cite{Petrilli2020, williamson2020, killerby2020}, cancer \cite{Petrilli2020, williamson2020, robilotti2020determinants}, diabetes  
\cite{Petrilli2020, williamson2020, killerby2020, diabetes},  obesity \cite{Petrilli2020, williamson2020, killerby2020, bello-chavolla_obesity, statcan2017obesity}, chronic kidney disease (CKD) \cite{Petrilli2020, williamson2020, killerby2020, AroraE417}, immuno-suppressed conditions \cite{williamson2020} and smoking \cite{statscansmoking, WHOsmoking}.
National prevalence estimates were extracted based on age group ($<$10, 11-20, 21-30, 31-40, 41-50, 51-60, 61-70, 71-80 and $>$80 years of age) and sex.
 
We determined conditional probability of developing COVID-19 in ABM based on symptoms and risk factors associated with COVID-19 in published literature.
A mathematical modelling study of the epidemic with Canadian-specific estimates~\cite{Davies_NatureMedicine2020} was used to model COVID-19 susceptibility in the pediatric population of the simulator.
 
\section{Inoculum \& Effective Viral Load}
\label{viral-load}
We model inoculum, the amount of virus transmitted during an exposure event, as a random variable uniformly distributed between $0.$ and $1.$. 
The magnitude of inoculum is used to determine the type and severity of symptoms. 

\begin{figure}[thp!]
\centering
\label{fig:effective_viral_load}
\begin{tikzpicture}[latent/.style={draw, circle, inner sep=0pt, thick, minimum size=20pt}, observed/.style={latent, fill=gray!50}, factor/.style={fill=black, rectangle, inner sep=0pt, minimum size=10pt}, small latent/.style={latent, minimum size=10pt}, small observed/.style={observed, minimum size=10pt}, edge/.style={thick, -{latex}}, small factor/.style={factor, minimum size=4pt}, x=20pt, y=20pt, double/.style={{latex}-{latex}, shorten >=1pt, shorten <=1pt}]
    \draw[fill=Green!30, draw=none] (0, 0) rectangle (1.5, 5);
    \draw[fill=Green!30, draw=none] (1.5, 0) rectangle (2.75, 5);
    \draw[fill=Orange!30, draw=none] (2.75, 0) rectangle (4, 5);
    \draw[fill=red!30, draw=none] (4, 0) rectangle (7, 5);
    \draw[fill=Orange!30, draw=none] (7, 0) rectangle (9, 5);
    \draw[fill=Yellow!30, draw=none] (9, 0) rectangle (11, 5);
    \draw[-{latex}, thick] (-0.5, 0) -- (12, 0);
    \draw[-{latex}, thick] (0, -0.5) -- (0, 5);
    \node[rotate=90] at (-0.5, 2.5) {Effective Viral Load};
    \node[] at (12, 0.5) {Time};
    
    \draw[ultra thick] (0, 0) -- (1, 0) -- (2.5, 4.5) -- (3.5, 3) -- (7, 3) -- (11, 0);
     \node[font=\scriptsize, align=center] at (1.375, 4.5) {Incubation};
     \node[font=\scriptsize, align=center] at (3.375, 4.5) {Onset};
    \node[font=\scriptsize, align=center] at (5.5, 4.5) {Plateau};
    \node[font=\scriptsize, align=center] at (8, 4.5) {Post \\plateau 1};
    \node[font=\scriptsize, align=center] at (10, 4.5) {Post \\plateau 2};
\end{tikzpicture}
\caption{
Schematic showing the viral load curve, and associated phases of symptoms with severity indicators: infectiousness onset occurs on average 2.5 days after exposure, viral load peaks 0.7 days before symptom onset, which occurs an average of 5 \textbf{incubation days} after exposure. Symptoms are most severe after viral load peak and symptom onset, when the virus has had time to infect many cells. Recovery takes on average 14 days from symptom onset.}
\end{figure}
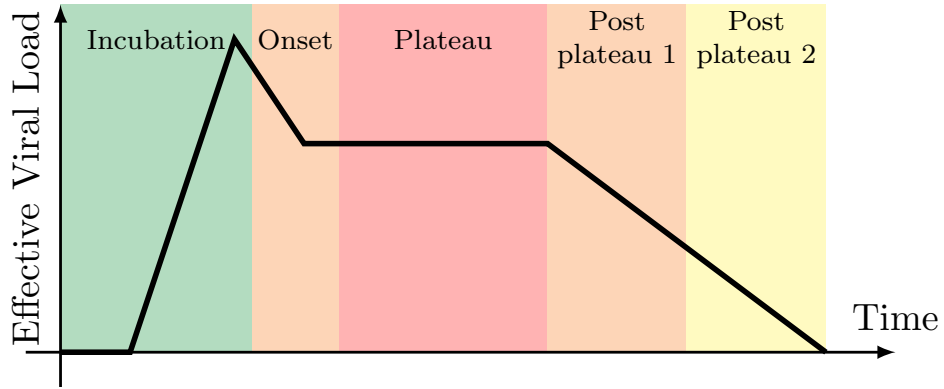

\begin{figure}
    \centering
    \includegraphics[width=\textwidth]{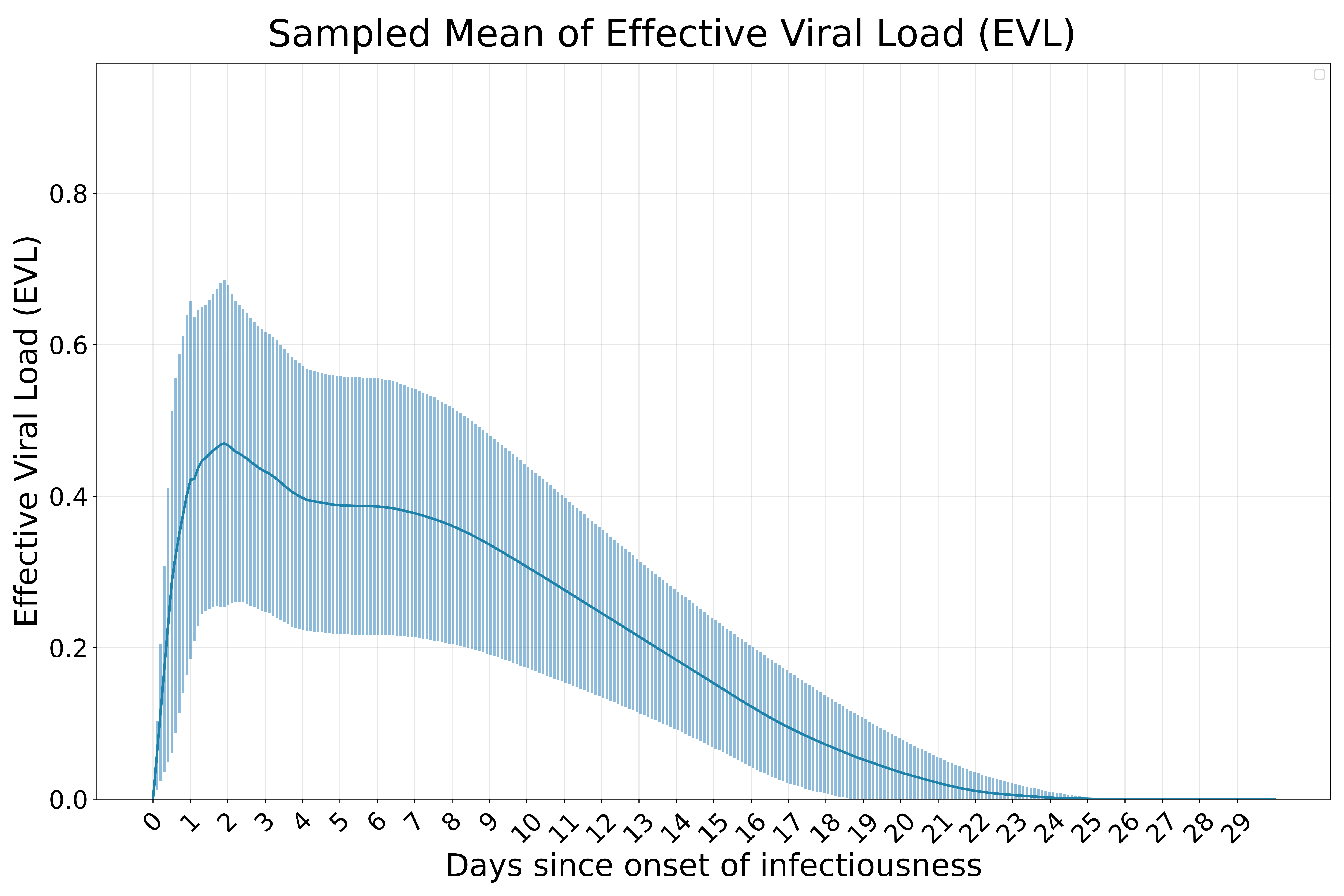}
    \caption{Mean $\pm$ 95\% C.I of sampled effective viral load of 40 agents. Note that the individual viral load curves are sampled as discussed in Appendix~\ref{viral-load}.}
    \label{fig:sim-viral-load}
\end{figure}

We sample parameters for a piece-wise linear model of what we call \textbf{effective viral load} (EVL)\footnote{Viral load is the number of actual viral RNA in a person; we model a number between 0-1 which could be converted to an actual viral load via multiplying by the maximum amount of viral RNA detectable by a given test.}.
We think of EVL as a piecewise linear function, attributes of which are sampled for each individual separately.
This approximation follows empirical studies on viral load progression~\cite{to2020temporal, lauer2020incubation}.
Figure~\ref{fig:sim-viral-load} is the mean of sampled effective viral load curve. 

\section{Testing \& Quarantine} \label{app:testing}%
We model RT-PCR test allocation as a priority queue. 
When an agent experiences symptoms, or is advised by a contact tracing application to seek a test, this agent is added to the queue. 
For different symptom severities, we assign a different probability of seeking a test. 
At some points during the pandemic, Montreal experienced a restricted testing capacity. 
In order to model that setting, we limit the daily testing capacity to a proportion of the population (generally set to 0.1\%).

To determine which people get tested under such a restriction, we allocate tests to people with more severe symptoms and app-based recommendations. 
After a test has been conducted, there is a delay before results are returned (2 days in our experiments). 
During this period, the individual who is being tested is isolated, and if the test returns positive then an additional 12 days of isolation is recommended. 
Any quarantine which is applied to the person receiving the test is also applied to other members of their household.

\section{Hospitalization} \label{app:hospitalization} %
We adopt a simple model of hospitalization and interaction within them. 
To simulate post-lockdown scenarios, we assumed no infections in hospitals.
We adopt a probabilistic model of hospitalization where likelihood of being admitted to the hospital or ICU depends on symptom severity and age. 
Mortality rates are conditioned on age-group following data from Quebec public health (date?) \cite{quebec2020publichealth}. 
The duration of a hospitalization, likelihood of requiring critical care, and mortality rate given critical care by age follow nationally conducted surveys available publicly\footnote{\url{https://health-infobase.canada.ca/COVID-19/epidemiological-summary-COVID-19-cases.html\#a8}}. 
The number of hospitals is defined in relation to the population, using the same ratio of hospitals to people as are found in Quebec (1.99 hospitals per 100,000 people). 
The number of hospital beds per capita, and occupancy ratios are taken from \footnote{\url{https://www.cdc.gov/nchs/data/hus/2017/089.pdf}} , and the number of icu beds per capita and occupancy ratios are taken from \footnote{\url{https://www.sccm.org/Communications/Critical-Care-Statistics}}. 
Hospitals are staffed by doctors and nurses, who are modelled as people with a profession that requires they work at the hospital and have protected interactions with patients.

Literature on the link between underlying medical conditions and COVID-19 encompasses risk of being hospitalized due to COVID-19, risk of severe complications (e.g. mechanical ventilation) and risk of mortality. 
Pre-existing medical conditions were used to model: 1) hospitalizations and deaths outcomes and 2) conditional probability of symptoms. %
To model hospitalizations and deaths in the population, we used risk ratio estimates from studies focusing on risk of hospitalisation and risk of death as outcomes. 
Risk ratios adjusted for other individual characteristics were preferred over crude estimates.
  
To inform the hospitalisation and death outcome model from the simulator, we selected the following risks ratios: diagnosis of heart disease, 1.17 \cite{williamson2020}, stroke history, 2.16 \cite{williamson2020}, asthma with recent oral corticosteroid use, 1.13 \cite{williamson2020}, COPD, 1.08 \cite{Petrilli2020}, %
cancer (excluding haematological malignancy), 1.72 \cite{williamson2020}, diabetes, 2.24 \cite{Petrilli2020}, obesity stages 1 and 2 (body mass index (BMI) = 30-39.9 kg/m$^2$), 1.8, obesity stage 3 (BMI $>$40 kg/m$^2$), 2.45 \cite{Petrilli2020}, CKD, 2.60 \cite{Petrilli2020}, immuno-suppressed because of asplenia, 1.34 and because of immunosuppressive conditions (excluding asplenia and haematological malignancy), 1.70 \citep{williamson2020}. Given the uncertainty on the association between smoking and COVID-19 prognosis   \citep{WHOsmoking}, we did not consider this risk factor in the simulator.

\section{Contact Patterns} \label{app:mobility} %
In this section we describe contact patterns in pre-pandemic situation.
Scenarios of lockdown and contact patterns in intermediate behavior levels are a modification by a factor $\gamma_n$ as discussed in the section~\ref{sec:abm}.

\subsection{Empirical Matrices} \label{app:contact-derivation}
We use empirically derived matrices in 2017 from~\cite{Prem2017ProjectingSC} for Canada that we further project on to Montreal's demographical structure. 
Projection of country-wide matrix to a regional matrix is done via method described in~\cite{arregui2018projecting}.
However, ABM can be configured to bypass the step of regional projection of contact matrices. 
Given a discrepancy between population wide mean daily contacts inferred from projected matrices and Montreal's number of contacts reported in a 2020 survey~\cite{brisson2020}, we scale the projected matrices appropriately. 
We ran 12 simulations with no infected agent i.e. $\alpha=0$ to observe the pre-pandemic contact patterns to yield simulated contact patterns in this section.
Additionally, the simulated contact patterns shown in this section are descaled with the same multiplicative factor that is used to scale the projected matrices.

\subsection{Dwelling Characteristics} \label{app:dwelling}
As discussed in the main section, agents are grouped into houses according to census data~\cite{statistics2016census}.
Thus, we simulate dwelling characteristics of the city of Montreal.
However, it can be configured easily for any other city by using appropriate parameter values. 
We consider house of sizes ranging from 1 to 5.
Age distribution of agents living solo also follows census data. 
Further, house sizes ranging from 2 to 5 consider three broad categories of dwelling characteristics - (a) couple with $x$ kids, (b) single parent with $x$ kids, (c) random allocation, where $x$ represents number of kids required to complete the house size. 
For example, for a house with single parent and of size 5, $x=4$. 
The distribution of these characteristics also follow from census data.
Finally, we also consider senior residencies where a proportion of agents above age 65 live. 
We inform this proportion from the census data as well\footnote{\url{https://www12.statcan.gc.ca/census-recensement/2016/dp-pd/covid19/table2-eng.cfm?geo=A0002&S=1\&O=A}}.

\begin{figure}[!htp]
    \centering
    \includegraphics[width=\textwidth]{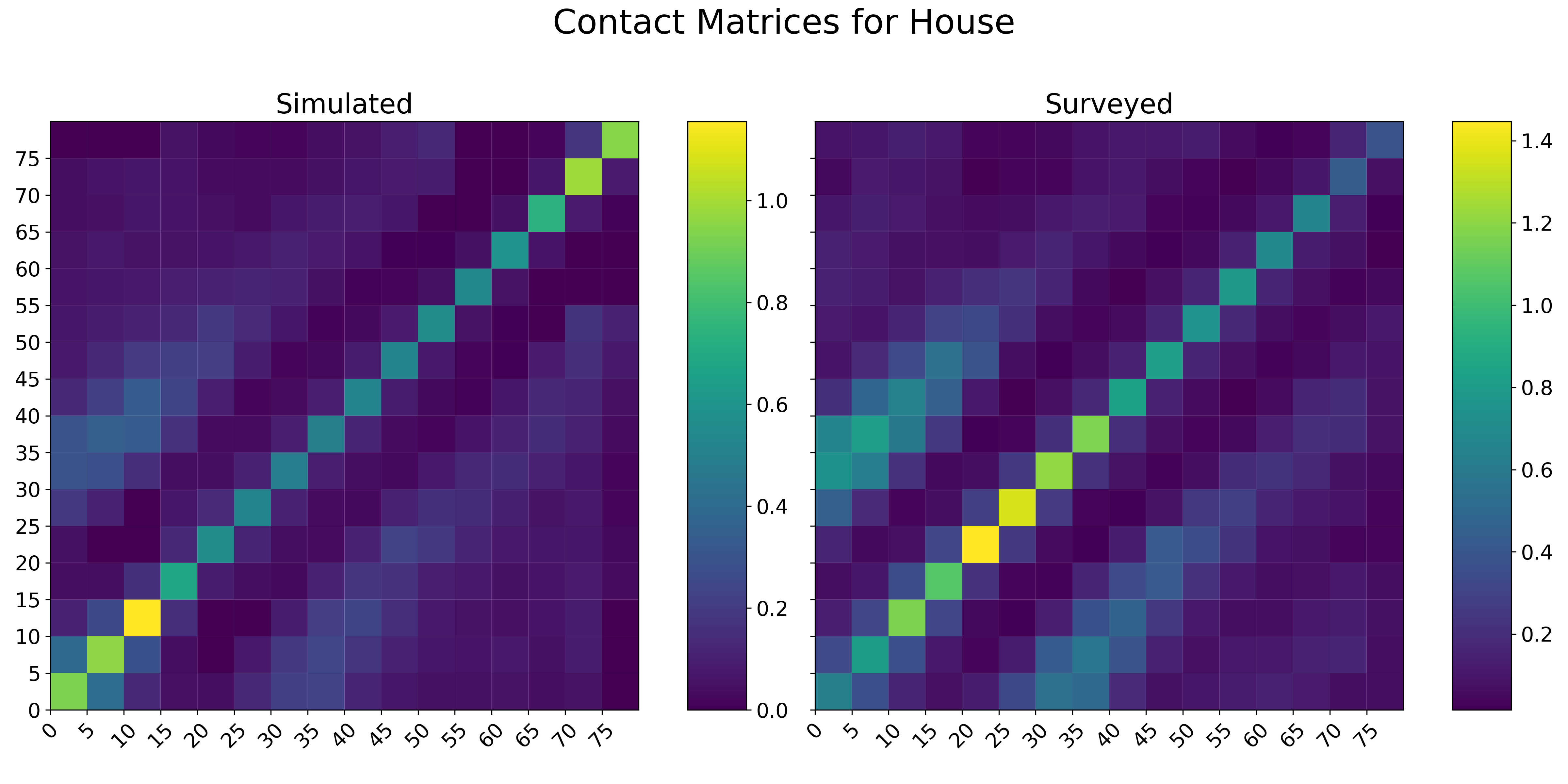}
    \caption{Housing allocation of agents is calibrated to yield a contact pattern (left) similar to the empirically derived household contact matrices (right) described in Appendix~\ref{app:contact-derivation}. We explicitly model older adults living in assisted care resulting in oversampling of contacts in that age group. We discuss it further in the Appendix~\ref{app:dwelling}}
    \label{fig:house-contacts}
\end{figure}

As a result of housing allocation discussed above, we yield a contact pattern as shown in Figure~\ref{fig:house-contacts}.
We make two observations (a) there is an oversampling of contacts towards the older age groups: It is because older agents grouped in collectives like senior residencies are modelled explicitly. This choice was motivated from~\cite{zimmerman2020need} which suggests inclusion of collectives in proper response to the COVID-19 pandemic. 
(b) a slight discrepancy we observe in the intensity of the main diagonal is due to insufficient social gatherings at households.

\subsection{Workplace Allocation} 
\begin{figure}[!htp]
    \centering
    \includegraphics[width=\textwidth]{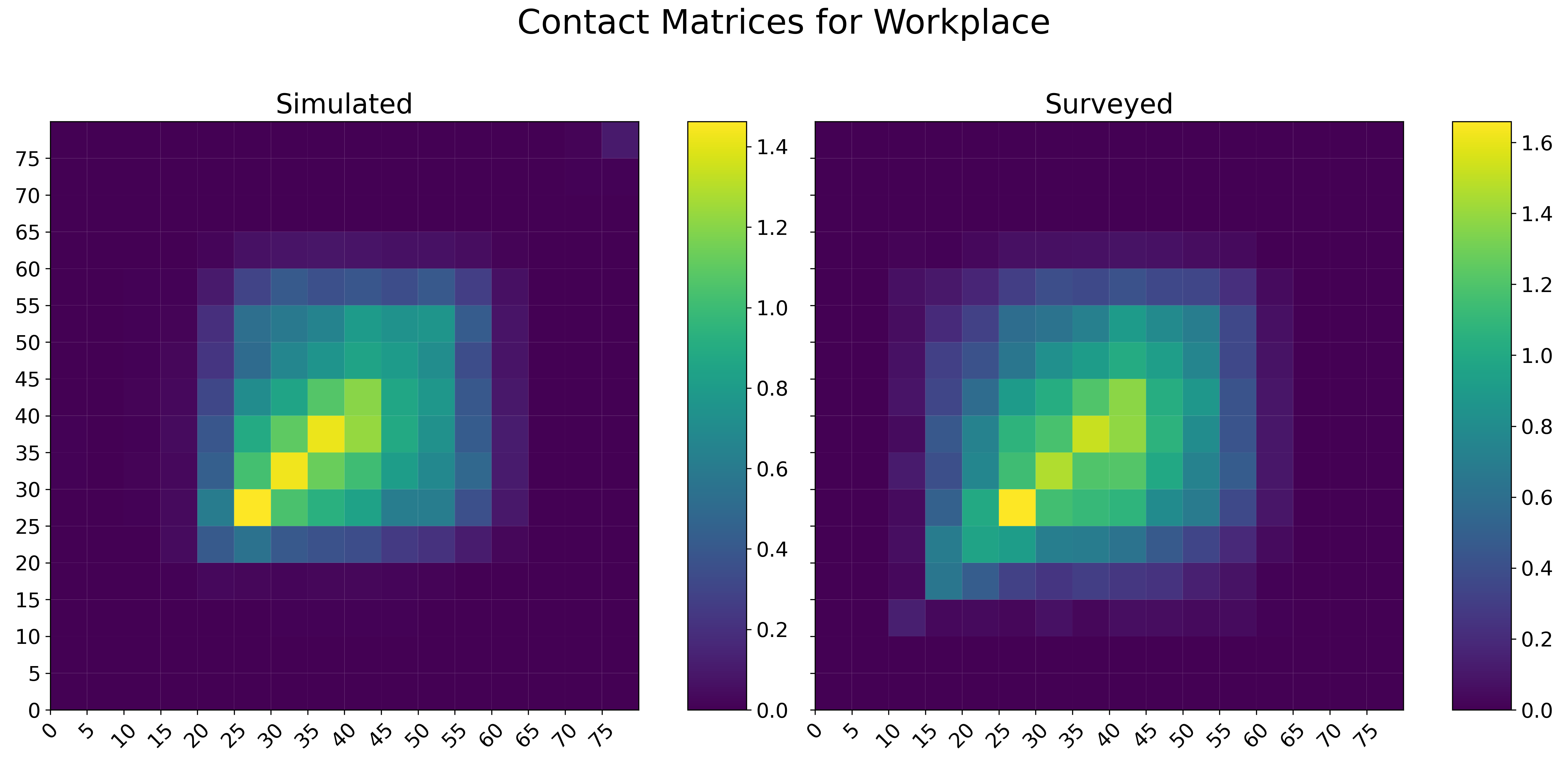}
    \caption{Simulated contact pattern (left) at workplaces follow empirically derived matrix (right) projected onto regional demographical structure as described in Appendix~\ref{app:contact-derivation}}
    \label{fig:workplace-contacts}
\end{figure}

\begin{figure}[!htp]
    \centering
    \includegraphics[width=\textwidth]{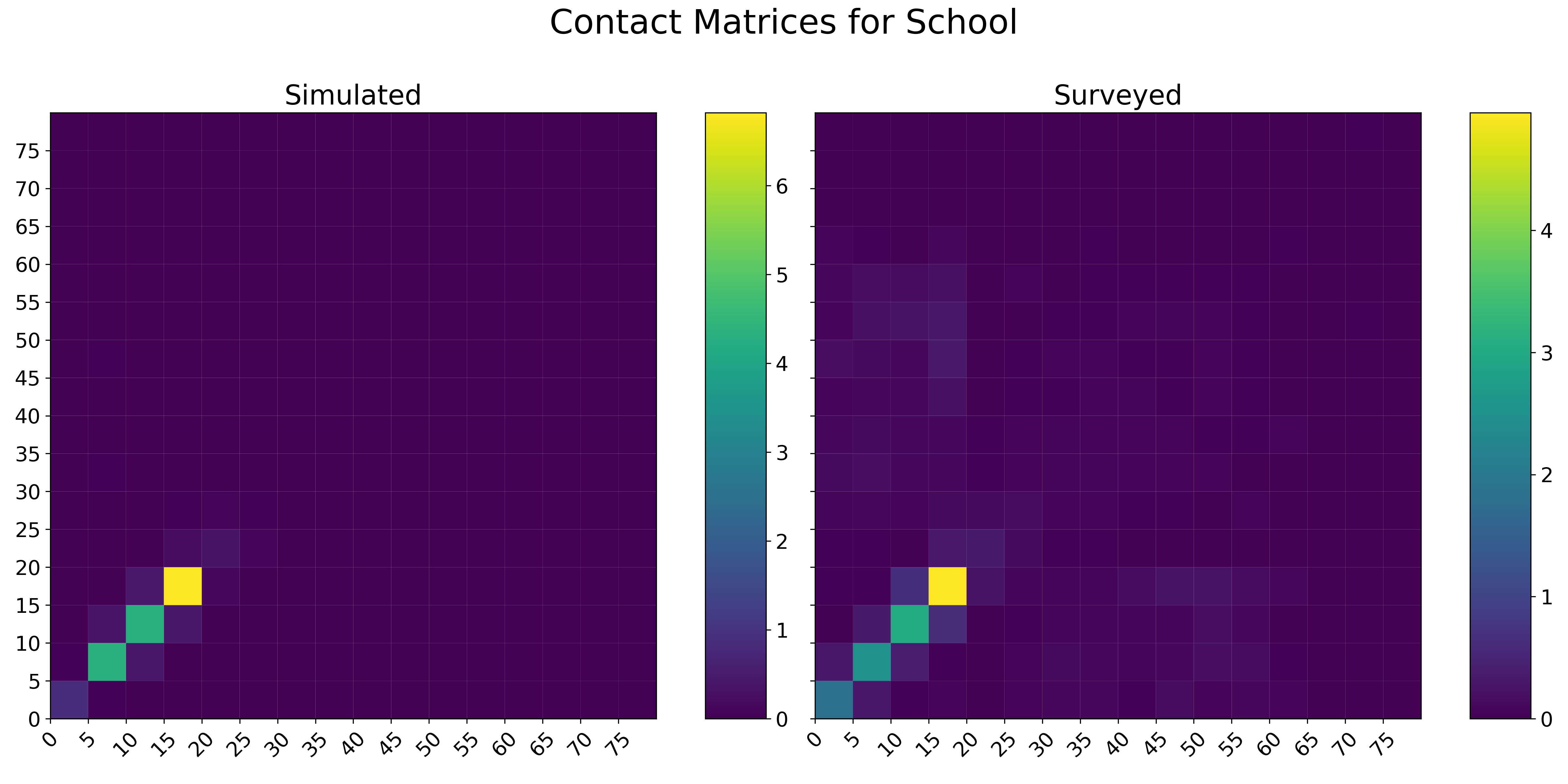}
    \caption{Simulated contact pattern at schools (left) yield a similar pattern to empirically derived matrix (right) as described in Appendix~\ref{app:contact-derivation}.}
    \label{fig:school-contacts}
\end{figure}

We consider an age-dependent workplace allocation such that agents in each age group have a probability of attending a school or a workplace.
We consider schools for the following age groups (a$^*$) 2-4 years old (y.o) (b) 4-5 y.o (c) 5-12 y.o (d) 12-17 y.o (e$^*$) 17-19 y.o: (f$^*$) 19-24 y.o (g$^*$) 25-29 y.o, where $^*$ marks the age group in which only a fraction (informed by census data) of agent population was allocated schools.
Further, we assume 100\% employment so that all agents older than 17 y.o and younger than 65 y.o. were allocated a workplace. 
Agents in senior residencies are allocated a common room as their workplace where they get together during working hours.
Such allocation of younger agents to schools give a contact pattern as shown in Figure~\ref{fig:school-contacts}, while a workplace allocation of adults yield a contact pattern shown in Figure~\ref{fig:workplace-contacts}.


\subsection{Random (Other) Locations}
\begin{figure}[!htp]
    \centering
    \includegraphics[width=\textwidth]{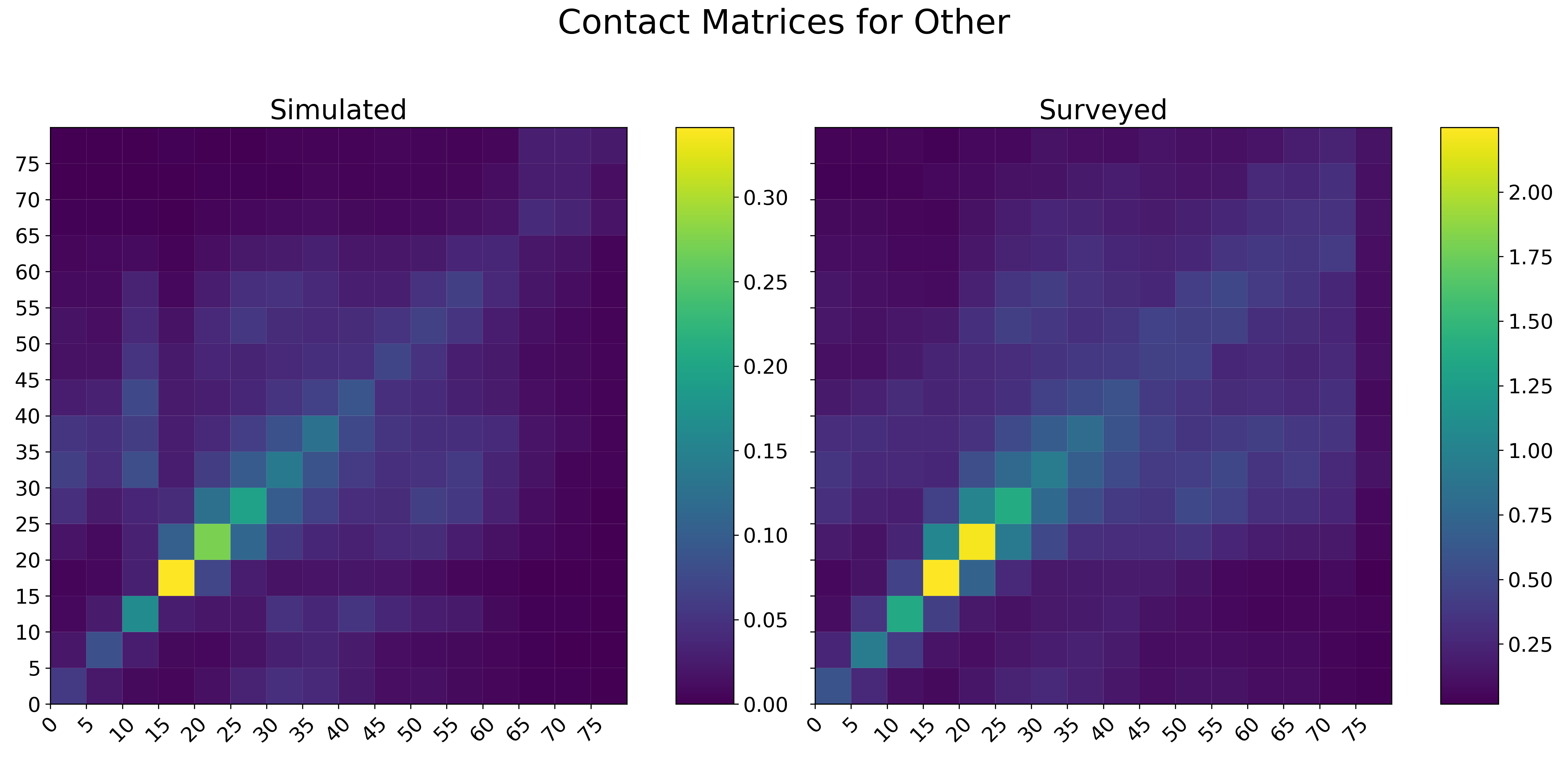}
    \caption{Simulated contact pattern at "Other" locations (left) yield a similar pattern to empirically derived matrix (right) as described in Appendix~\ref{app:contact-derivation}. Agents spend relatively shorter duration of time at these locations as compared to house or workplaces. Examples of such locations include parks, restaurants, and grocery stores. We observe a slight undersampling of such contacts due to improper estimates of time spent at such activities.}
    \label{fig:other-contacts}
\end{figure}

To model infections at locations other than workplace and house, we consider locations where agents remain for a relatively shorter duration as compared to house and workplace. 
Specifically, we model interactions at locations like restaurants, grocery stores, and parks.
Note that this category of locations is also termed as ``other" in ~\cite{Prem2017ProjectingSC}.
Further, as the mean number of contacts at house are greater than the number of residents, it was important to consider socializing activities organized at houses.
To do this, we maintain a pool of agents that an agent interacts with, and bring them together for a social activity at either a restaurant or at the agent's house.
We discuss scheduling of these activities next. 
A contact pattern resulting at these random locations is shown in Figure~\ref{fig:other-contacts}.

\subsection{Activity Scheduling \& Mobility Pattern}
We consider \emph{adult supervision} for agents below 14 y.o i.e. except for when agent goes to school, at least one adult agent (older than 14 y.o) has to be present all the time. 
Thus, we pre-plan the schedule of agents older than 14 y.o at the start of the simulation, and plan the schedule of agents younger than 14 y.o during the simulation. 
Planning the schedule takes into account workplace opening hours as well as regularly scheduled activities like social gatherings\footnote{\url{https://www150.statcan.gc.ca/n1/pub/89-652-x/89-652-x2014006-eng.htm}}, exercising, and grocery shopping\footnote{\url{https://www.statista.com/statistics/944310/how-often-consumers-visit-food-stores-canada/}}. 
Thus, an agent who has gone to a grocery store or a restaurant on one day will be less likely to go again during that same week, and so on. 
The schedule additionally depends on the day of the week. 
For instance, agents with school as workplaces are scheduled to be at school on weekdays, whereas most of their time will be spent at home on weekends.

\begin{figure}[!htp]
    \centering
    \includegraphics[width=\textwidth]{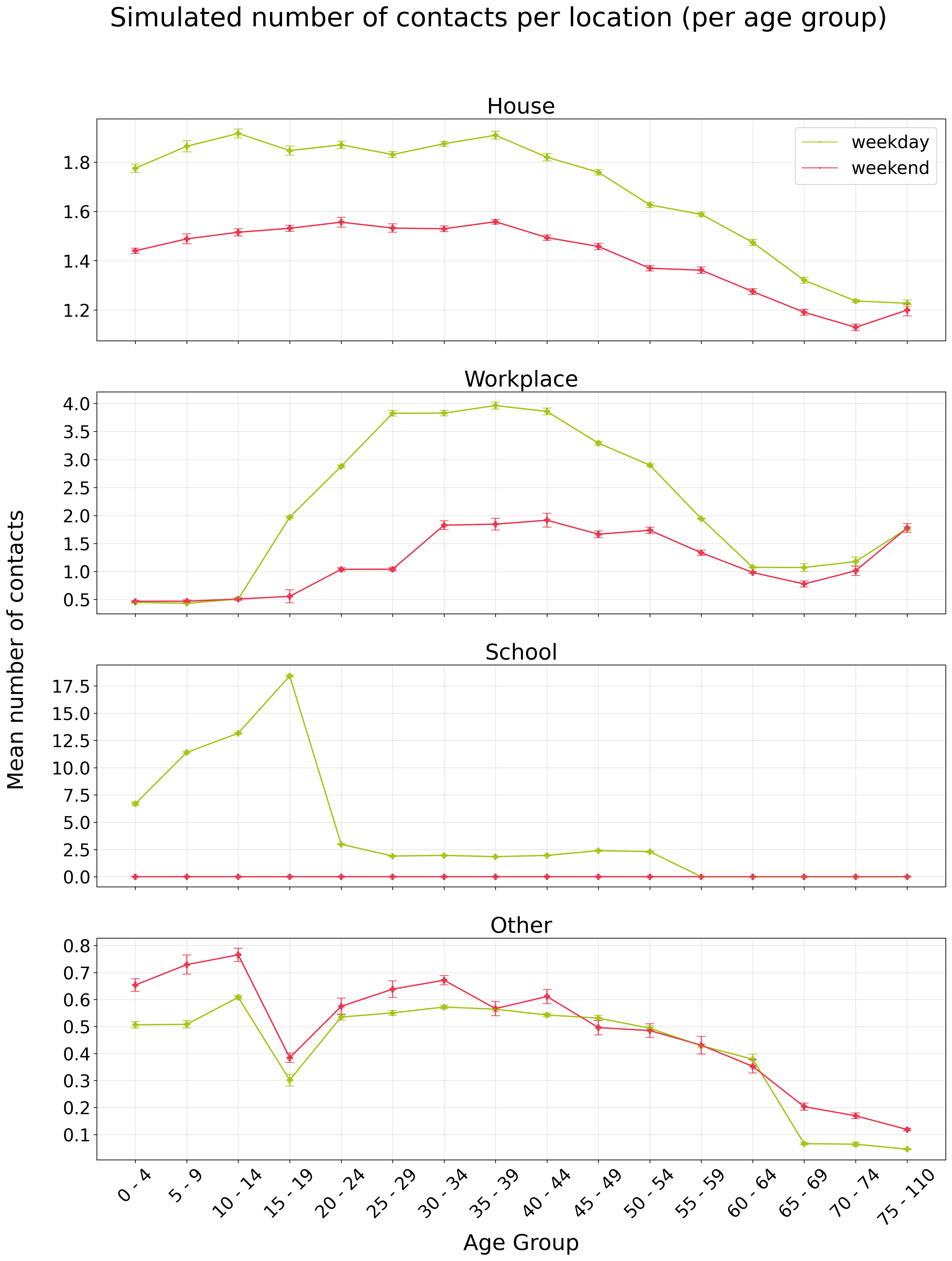}
    \caption{Simulated mean daily contacts on weekdays and weekends broken down by age groups. Agent activities are scheduled such that the mean number of contacts on work and non-work days follow surveyed data as reported in~\cite{klepac2020contacts, mossong2008social} }
    \label{fig:mean-daily-contacts}
\end{figure}

On the day of activity, however, these activities might stand cancel due to sickness, quarantining requirements, or hospitalization.
In these situations, location of activity is appropriately changed for a required duration. 
At the same time, if an agent requiring adult supervision is sick, has to quarantine, or is hospitalized, an adult from the same house has to cancel the activity to stay with the agent.
Of note, agent’s mobility i.e. presence in locations other than house is reduced when they are experiencing symptoms (to a degree proportional to symptom severity).
Note that we do not change the schedule unless the agent is quarantining i.e. normal mobility is maintained all the time unless the agent is put in the level $n$.
Figure~\ref{fig:mean-daily-contacts} show a breakdown of contacts at each location on weekdays and weekends.

\subsection{Sampling of Contacts}
\begin{figure}[!htp]
    \centering
    \includegraphics[width=\textwidth]{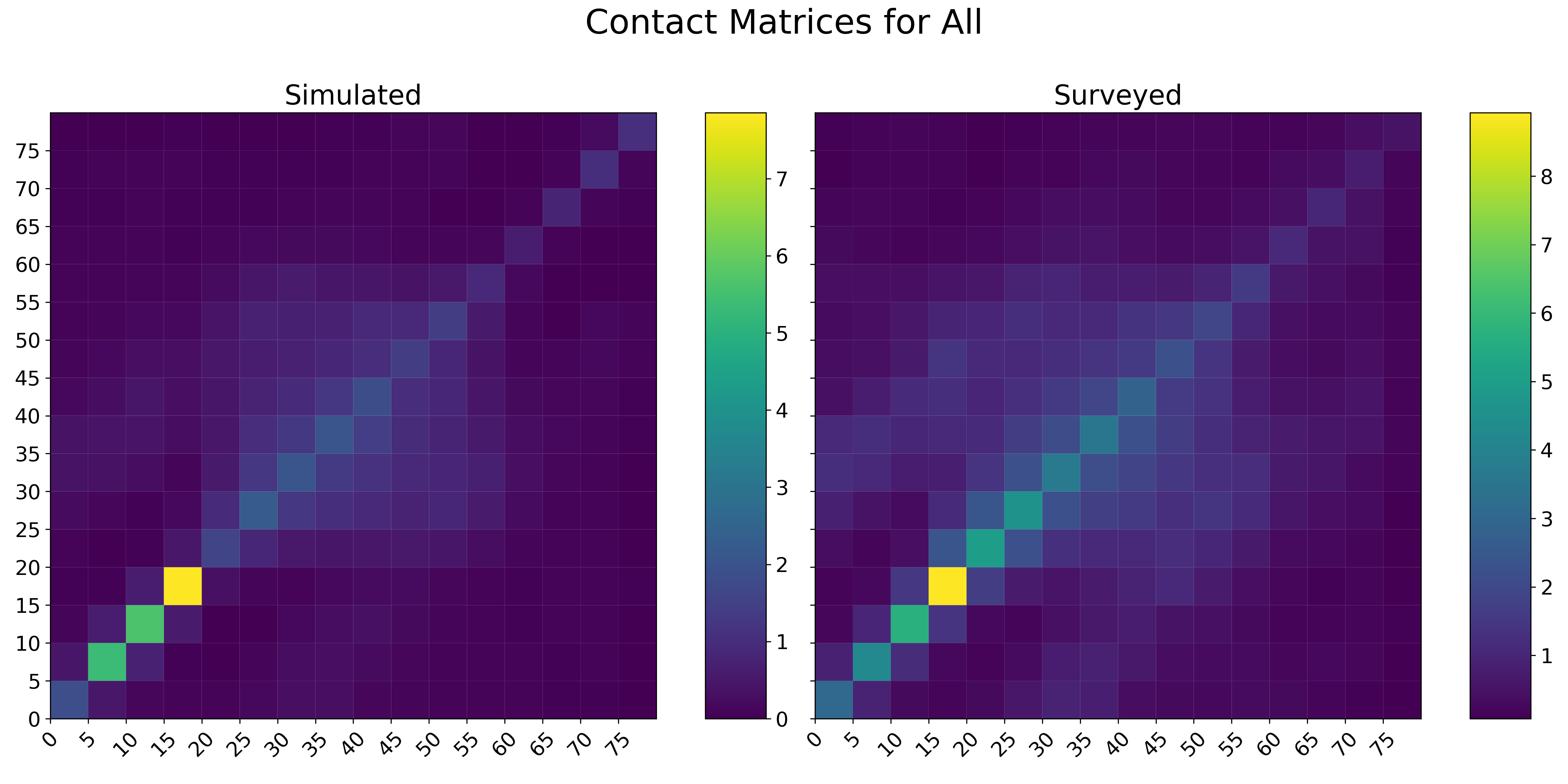}
    \caption{Overall simulated contact pattern (left) yield a similar pattern to empirically derived matrix (right) as described in Appendix~\ref{app:contact-derivation}.}
    \label{fig:all-contacts}
\end{figure}

We implement age-stratified contact sampling as informed from empirically derived contact matrices as described in Appendix~\ref{app:contact-derivation}. 
Specifically, for each agent we draw number of contacts as per the location-specific age-stratified number of contacts obtained from the contact matrices.
We use a negative binomial distribution~\cite{wiki:nb} to draw number of contacts.
Further, we use these matrices to infer probability of interaction with other agents in each age group, thereby, implementing location dependent assortativity in interactions.
We call these interactions as \emph{encounter}.
Finally, we also draw amount of time spent in each encounter as per the survey conducted in California~\cite{zagheni2008using} standardized to the demographics of Montreal.
Thus, we obtain an aggregated contact pattern as shown in Figure~\ref{fig:all-contacts}. 

\section{Simulator Customization}
\label{customizing-the-sim}
We briefly describe the required steps for customization of the simulator. Location-specific demographic and contact data may be modified simply by adding a new configuration file to the configs folder.
Configuration files are written using YAML, a human friendly data serialization standard. 
Essentially, these files contain key-value pairs.
The values for a new region must be specified and contained in the new configuration file. 
Examples of modifications required to model a new region include: population-level distribution of age, housing distribution i.e. number of houses from size 1 to 5, occupation characteristics including age for kids to go to schools, retirement age, etc. and, finally, how often people go out to stores, socialize, 

\section{Privacy, Security, and Phone technology} \label{app:privacy-messaging}
We provide details of the messaging protocol that uses Bluetooth signals to exchange tokens. The privacy protocol which provides anonymous message exchange between phones which have exchanged these tokens is covered in more depth in another paper \footnote{full citation to be provided in the camera-ready version}.
In the context of the current work, we focus on the description of how we model the Bluetooth communication range of phones and on the description of the clustering algorithm we use to prepare received messages for input to a risk prediction model.

\subsection*{Bluetooth Details}\label{app:bluetooth}
Given a ground-truth distance and duration for our encounter, we want to determine if two app-users should exchange encounter messages. We only exchange encounter messages if the perceived distance is below 2 meters and the duration greater than 15 minutes. To compute this, we use a naive model of the Bluetooth noise which uses a per-person noise and a per-encounter noise. Each smartphone has a different noise sampled uniformly between 0 and 1. Each encounter takes the mean of this value across both users. We then apply a relative offset to the real encounter distance by multiplying the combined user noise and a uniform random variable centered at 0 with range $[0.5, -0.5]$. The magnitude of the distance offset is up to 2 metres when the real distance is 2 metres, and up to 0.5 metres when the ground truth distance is 1 metre.

\subsection*{Encounter Messages}\label{app:enc-messages}
An encounter message $m^{enc}$ is composed of the day the encounter occurred $d$ and the sender's quantized risk at the time of the encounter $r_d$. The formal definition of a risk message is: 
$$ m^{enc} = (r_d, d) \in \mathcal{M}^{enc} $$
An app user (say ``Alice'') receives new encounter messages four times per day in batches; we call the set of new encounter messages on day $d$, $\mathcal{M}^{new}$. The risk messages which Alice has already clustered are noted $\mathcal{M}^{enc}$ (which is an empty set when Alice first installs the app). 
It is useful to think of encounter messages as database records. Alice inserts records into her database whenever she encounters other app users (``Bobs'') with their current risk estimate. Update messages are like database update statements, and are used to change the risk values in old encounter messages.

\subsection*{Risk Update Messages}\label{app:update-messages}
If a user (say ``Bob'') had many encounters with Alice, and Bob subsequently receives a positive test result, becomes sick (reporting symptoms), or otherwise re-evaluates his risk, then Bob will send risk update messages to Alice. Similar to encounter messages, these risk update messages may be sent through the server up to four times per day. More specifically, if Bob updates his estimate of his risk on some previous day $d_{old}$, then Bob will construct a risk update message for every encounter message that he had on $d_{old}$ and these messages are sent to the users with whom Bob had a contact. If this update message is sent on day $d$, it is composed of the current day $d$, his new risk $r_{new}$, his previous risk estimate for that day $r_{old}$ and the day of the encounter $d_{old}$. Therefore, the formal definition of a risk update message is:
$ m^{update} = (r_{new}, d, r_{old}, d_{old}) \in \mathcal{M}^{updates} $

\subsection*{Encounter and Update Message Clustering}\label{app:clustering}
By virtue of the strong privacy protocol in place, we are not able to create message clusters that correspond the true chains of contacts for the encountered users. The only information we have to create clusters is the day and risk level of the encounters. Our hypothesis is that a user's contact patterns can provide a rough idea on the number of individual people they encounter on each day.

All encounter messages received on a given day with a similar risk level are put into the same cluster. Given that there are 16 risk levels, it is only possible to create 16 new clusters on a given day, unless the user receives update messages. Every update messages is created for exactly one encounter message. If there are fewer update messages for a given day / risk combination than there are encounter messages for that day / risk, then we split the existing cluster for that day / risk into two, with one cluster containing newly updated encounter messages, and another containing the rest of the messages.

We do not claim that this clustering algorithm is optimal, that the input format to the neural network is optimal, or that this messaging protocol is optimal. There exists an interesting trade-off between privacy and risk precision which we hope will be explored in future work.

\section{Transmission} %

\textbf{Agent-Agent Transmission} All encounters are assumed to be at a distance equal or less than two meters. 
Transmission events can only take place between infectious and susceptible individuals.
Although the ABM models all encounters taking place between individuals, a susceptible individual is only considered exposed (i.e. the model considers that an effective contact has taken place) when the encounter lasts a minimum of 15 minutes and bernoulli distribution with probability determined via Equation~\ref{eq:prob_transmission} gives 1. 
The likelihood of viral transmission during a given effective contact is proportional to the time duration of the encounter, and depends broadly on characteristics of both the infected and susceptible individual. 

We use the transmission model as explained in~\cite{ferretti2020quantifying}.
Thus, encounters taking place in certain locations (i.e. senior residencies and households) are also inherently more prone to result in a transmission event, all other factors being equal.
This is modeled via $B_n$.
At the same time, characteristics affecting infectiousness (of the infected individual) include progression of the disease (i.e. effective viral load) ($EVL$), whether (and to what extent) they are symptomatic ($A_s$).
Susceptibility of the exposed individual depends notably on age ($S_a$). 

\textbf{Environment-Agent Transmission} Please note, that the probability of environmental transmission is not supported by any published study, and we consider 0 environmental transmissions in the experiments in this paper.
However, ABM can be configured to simulate environmental transmissions. 
Empirical estimates of such transmissions stands at 10\% as per~\cite{ferretti2020quantifying}, therefore, we consider a transmission model that models environmental infections in pre-pandemic scenario. 
Given the lack of such estimation in post-lockdown scenario, we do not consider any such transmissions. 

We model these transmissions by considering a linearly time-decaying probability of location being \emph{contaminated} which is triggered by the presence of an infectious individual.
Initial magnitude of contamination is dependent on the agent's current phase of the disease.
Further, the duration of such contamination is informed from experimental study~\cite{suman2020sustainability}, which lists surfaces and the duration for which virus survives on them.
We consider an experimental environment transmission model that estimates probability of infecting agent as proportional to (a) contamination strength of the location, and (b) susceptibility of an agent.
The proportionality factor being the \emph{environmental transmission control knob} which lets us model the disease spread as per the observed data.

\section{Heuristic-FCT} \label{app:heuristic-algo}

We denote $S_{HIGH}$ as the row indices in $\mathbf{S}_d^i \in \{0, 1\}^{N_{symptoms} \times d_{max}}$ such that the symptom at those indices are highly informative symptoms (see Sec. \ref{app:informative-symptoms}). 
Similarly, we obtain sets of indices $S_{{\rm MODERATE}}$ and $S_{{\rm MILD}}$. 

We used $n+1=4$ recommendation levels, thereby allowing an agent behavior level to be either of $\{1, 2, 3, 4\}$.
 We also denote the set of days $\{d_1, d_2, ... d_{max}\}$ as $D$.
The {\em Heuristic-FCT} algorithm is implemented as Algorithm~\ref{alg:hct}. We next describe each function of the \textit{Heuristic-FCT}.

\begin{enumerate}
   \item \textbf{Compute Risk} calls each of the below functions in order to generate a risk history over the last $d_{max}$ days and a current recommendation level. It takes in symptom, test, and risk message data, then takes the element-wise maximum over the risk histories associated with these inputs. \textit{Apply Negative Test} and \textit{Handle Recovery} modify the merged risk history or return an alternative (respectively) to this risk history.
    \item \textbf{Test Results Compute Risk} covers two cases. A positive test result within the last $d_{max}$ days sets the risk over the past $d_{max}$ days to $r_{{\rm MAX}}$ and $\zeta^i_d$ to 3.
    \item \textbf{Symptoms Compute Risk} determines the risk and recommendation levels due to symptoms. We group symptoms by how informative they are of COVID-19. Highly informative symptoms experienced within the last $d - d_{max}/2$ days sets high risk levels into $\textbf{r}^i_{d:d-d_{max}/2}$, and $\zeta^i_d$ to 3. Moderately informative symptoms yield moderate risk levels, and so on.
   \item \textbf{Risk Messages Compute Risk} converts received risk messages into a risk history and recommendation level. We group risk messages into high, medium, and low risk groups, assigning corresponding risk levels to the $i$'s risk history between the day of receipt of that class of risk message and the current day.
   \item \textbf{Handle Recovery} returns the recent risk history to 0 in the absence of recent evidence of COVID-19. If there are no symptoms within the last 7 days, no recent risk messages, and no positive test results, then we set risk for the past $d_{max/2}$ days as 0.
   \item \textbf{Apply Negative Test} overwrites the element-wise maximum over the risk histories $r_d^i$ by assigning a $0$ risk to the agent around the days when a negative test was reported. Precisely, we denote by $W$ a time window, $W \in \mathcal{N}$ and $W < d_{max}$, such that we set a $0$ risk for the agent for $W/2$ days around the day on which negative test was reported. 
   
\end{enumerate}
\begin{algorithm}[htp]
	\caption{{\em Heuristic-FCT}}
	\label{alg:hct}
	\begin{algorithmic}[1]
	    \Function{TestResultsComputeRisk}{$\mathbf{T}^i_d$}
	    \State $\mathbf{r} \leftarrow \{0\}^{d_{max}}$ 
	    \State $\zeta \leftarrow 0$
	    \If{$\sum_{d' \in D}\mathbbm{1}_{\{\mathbf{T}^i_{d,d'} = +1\}} \geq 1 $}
	        \State $\mathbf{r}_{d:d-d_{max}} \leftarrow r_{{\rm MAX}}$
	        \State $\zeta \leftarrow 3 $
	   \EndIf 
	   \State \Return $\mathbf{r}$, $\zeta$
	 \par
	 \EndFunction 
	    \Function{SymptomsComputeRisk}{$\mathbf{S}^i_d$}
	    \State $\mathbf{r} \leftarrow \{0\}^{d_{max}}$ 
	    \State $\zeta \leftarrow 0$
	    \If{$\sum_{d' \in D, \; j \in {S_{{\rm HIGH}}}} \mathbf{S}^i_{d, \{j, d'\}} \geq 1$}
	        \State $\mathbf{r}_{d: d - d_{max}/2} \leftarrow r_{{\rm HIGH}}$ 
	        \State $\zeta \leftarrow 3$
	    \ElsIf{$\sum_{d' \in D,\; j \in {S_{{\rm MODERATE}}}} \mathbf{S}^i_{d, \{j, d'\}} \geq 1$}
	        \State $\mathbf{r}_{d: d - d_{max}/2} \leftarrow r_{{\rm MODERATE}}$ 
	        \State $\zeta \leftarrow 2$
	    \ElsIf{$\sum_{d' \in D,\; j \in {S_{{\rm MILD}}}} \mathbf{S}^i_{d, \{j, d'\}} \geq 1$}
	    	\State $\mathbf{r}_{d: d - d_{max}/2} \leftarrow r_{{\rm MILD}}$ 
	    	\State $\zeta \leftarrow 1$
        \EndIf 
        \State \Return $\mathbf{r}$, $\zeta$
	    \EndFunction 
	    \par
	    \Function{RiskMessagesComputeRisk}{$M_{i, :}^:(:)$}
	    \State $\mathbf{r} \leftarrow \{0\}^{d_{max}}$ 
	    \State $\zeta \leftarrow 0$
	    \If{$\sum_{j \in \mathcal{N}(i), d' \in D, d''}\mathbbm{1}_{\{M_{i, j}^{d'}(d'') = r_{{\rm MAX}}\}} \geq 1$}
	        \State $\hat{d} \leftarrow \text{Earliest day of receiving } r_{{\rm MAX}}$
	        \State $\mathbf{r}_{d:\hat{d}} \leftarrow r_{{\rm MODERATE}}$
	        \State $\zeta \leftarrow 2$
        \ElsIf{$\sum_{j \in \mathcal{N}(i), d' \in D, d''}\mathbbm{1}_{\{M_{i, j}^{d'}(d'') = r_{{\rm HIGH}}\}} \geq 1$}
	        \State $\hat{d} \leftarrow \text{Earliest day of receiving } r_{{\rm HIGH}}$
	        \State $\mathbf{r}_{d:\hat{d}} \leftarrow r_{{\rm MILD}}$
	        \State $\zeta \leftarrow 1$
        \ElsIf{$\sum_{j \in \mathcal{N}(i), d' \in D, d''}\mathbbm{1}_{\{M_{i, j}^{d'}(d'') = r_{{\rm MODERATE}}\}} \geq 1$}
	        \State $\hat{d} \leftarrow \text{Earliest day of receiving } r_{{\rm MODERATE}}$
	        \State $\mathbf{r}_{d:\hat{d}} \leftarrow r_{{\rm MILD}}$
	        \State $\zeta \leftarrow 1$
	   \EndIf 
	   \State \Return $\mathbf{r}$, $\zeta$
	   \EndFunction 
      \algstore{hct}

	\end{algorithmic}
\end{algorithm}

\begin{algorithm}[htp]        
	\caption{{\em Heuristic-FCT}}
	\label{alg:hct2}
    \begin{algorithmic} [1]  
        \algrestore{hct}
	    \Function{HandleRecovery}{$\textbf{S}_d^i$, $\textbf{T}_d^i$, $M_{i, :}^:(:), \mathbf{r}^i_d$}
            \State ${\rm Rx} \leftarrow 1$
    	    \If{$\sum\mathbf{S}^i_{d, \{:, d:d-d_{max}/2\}} \geq 1$ or $\sum_{d' \in D}\mathbbm{1}_{\{\mathbf{T}^i_{d,d'} = +1\}} \geq 1 $}
    	        \State ${\rm Rx} \leftarrow 0$
            \EndIf
            \If{$\sum_{j \in \mathcal{N}(i),\; d' \in D,\; d'' \in \{d, d-1, ... d-7\}}\mathbbm{1}_{\{M_{i, j}^{d'}(d'') = r_{{\rm HIGH}}\}} \geq 1$}
                \State ${\rm Rx} \leftarrow 0$
    	   \ElsIf{$\sum_{j \in \mathcal{N}(i),\; d' \in D,\; d'' \in \{d, d-1, ...d-4\}}\mathbbm{1}_{\{M_{i, j}^{d'}(d'') = r_{{\rm MODERATE}}\}} \geq 1$}
                \State ${\rm Rx} \leftarrow 0$
    	   \ElsIf{$\sum_{j \in \mathcal{N}(i),\; d' \in D,\; d'' \in \{d, d-1\}}\mathbbm{1}_{\{M_{i, j}^{d'}(d'') = r_{{\rm MILD}}\}} \geq 1$}
                \State ${\rm Rx} \leftarrow 0$
    	    \EndIf 
    	   
    	   \If{${\rm Rx} = 1$}
    	    \State $\mathbf{r}^i_{d,d:d-d_{max}/2} \leftarrow 0$
    	   \EndIf
    	   
    	   \State \Return $\mathbf{r}^i_d$, ${\rm Rx}$
	   \EndFunction 
        
      \par
	  \Function{ApplyNegativeTest}{$\zeta_d^i, \mathbf{r}^i_d, \mathbf{T}^i_d, W$}
	    \State $d_n \leftarrow \text{day of the latest negative test}$
	    \State $\textbf{r}^i_{d,\;d_n - W/2\;:\;d_n + W/2} \leftarrow 0$
	    \If{$\textbf{r}^i_{d,d} = 0$}
	      \State $\zeta_d^i = 0$
	    \EndIf
	    \State \Return $\mathbf{r}^i_d$, $\zeta_d^i$
	  \EndFunction 
	  
      \par
      \Function{ComputeRisk}{$\mathbf{T}_d^i, \mathbf{S}_d^i, M_{i, :}^:(:), \mathbf{X}_i, \mathbf{r}_{d-1}^i$}
    	\State $W \leftarrow 8$
        \State $\mathbf{r}_t^i, \zeta_t^i \leftarrow \textsc{TestResultsComputeRisk}(\mathbf{T}_d^i)$
        \State $\mathbf{r}_s^i, \zeta_s^i \leftarrow \textsc{SymptomsComputeRisk}(\mathbf{S}_d^i)$
        \State $\mathbf{r}_m^i, \zeta_m^i \leftarrow \textsc{RiskMessagesComputeRisk}(M_{i, :}^:(:))$
        \State $\textbf{r}_r, {\rm Rx} \leftarrow \textsc{HandleRecovery}(\mathbf{S}_d^i$, $\mathbf{T}^i_d$, $M_{i,:}^:(:), \mathbf{r}_{d-1}^i)$
        \If{${\rm Rx} = 1$}
            \State \Return $\textbf{r}_r$, $0$
        \EndIf
        \State $\mathbf{r}_d \leftarrow \max(\mathbf{r}_t, \mathbf{r}_s, \mathbf{r}_m, \mathbf{r}_{d-1})$ \Comment{element-wise maximum}
        \State $\zeta_d^i \leftarrow \max(\zeta_t, \zeta_s, \zeta_m)$
        \If{$\sum_{d' \in D}\mathbbm{1}_{\{\mathbf{T}_{d,d'} = -1\}}  \geq 1 $}
        \State $\mathbf{r}^i_d, \zeta_d^i \leftarrow \textsc{ApplyNegativeTest}(\zeta_d^i, \mathbf{r}^i_d, \mathbf{T}^i_d, W)$
        \EndIf
        \State \Return $\mathbf{r}^i_d$, $\zeta^i_d$
      \EndFunction
	\end{algorithmic}
\end{algorithm}

\section{Symptoms} \label{app:symptoms}

\begin{figure}[ht]
    \centering
    \begin{adjustbox}{center}
    \resizebox{1.3\textwidth}{!}{\begin{tikzpicture}[text width=2cm, align=flush center, level 1/.style={sibling distance=7em},
level 2/.style={sibling distance=6em}, level distance=2cm, edge from parent path={
(\tikzparentnode) |-   %
($(\tikzparentnode)!0.5!(\tikzchildnode)$) -| %
(\tikzchildnode)}]
    \node {Symptoms}
        child { node {Fever \textbf{[SY-4]}}
          child { node {Chills \textbf{[SY-15]}} }
        }
        child { node {Gastro}
          child { node {Diarrhea \textbf{[SY-7]}} }
          child { node {Nausea / Vomiting \textbf{[SY-8]}} }
        }
        child { node {Sneezing \textbf{[SY-6]}} }
        child { node {Cough \textbf{[SY-2]}} }
        child { node {Fatigue \textbf{[SY-10]}}
          child { node {Unusual} }
          child { node {Hard time waking up \textbf{[SY-11]}} }
          child { node {Headache \textbf{[SY-9]}} }
          child { node {Confused \textbf{[SY-17]}} }
          child { node {Lost consciousness \textbf{[SY-18]}} }
        }
        child { node {Runny nose \textbf{[SY-14]}} }
        child { node {Sore throat \textbf{[SY-12]}} }
        child { node {Severe chest pain \textbf{[SY-16]}} }
        child { node {Trouble breathing \textbf{[SY-3]}}
          child { node {Light trouble breathing} }
          child { node {Mild trouble breathing} }
          child { node {Moderate trouble breathing} }
          child { node {Heavy trouble breathing} }
        }
        child { node {Loss of taste \textbf{[SY-5]}} }
        child { node {Aches \textbf{[SY-13]}} };
\end{tikzpicture}}
    \end{adjustbox}
    \caption{List of the symptoms in the simulator, with the corresponding feature IDs. The feature hierarchy is semantically meaningful -- if you have (SY-17, Confused) then you also have (SY-10, fatigue). We have constructed our symptom features this way to make the symptom reporting screens in the mobile application easier to use, in cooperation with a professional app-development company.  }
    \label{fig:list-symptoms}
\end{figure}
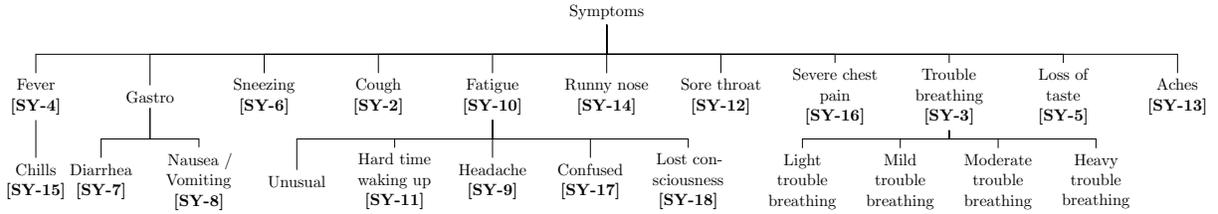

\paragraph{COVID-19 and other disease symptoms} %
Symptoms in the model include: Fever, Chills, Gastroenteritis (Gastro), Diarrhea, Nausea/vomiting, Sneezing, Cough, Fatigue, Hard time waking up, Headache, Confused, Loss of consciousness, Runny nose, Sore throat, Chest pain, Trouble breathing, Loss of taste/smell, and Aches, as shown in Figure \ref{fig:list-symptoms}. 
We represent symptoms as a set of diseases that are sampled on each day.

In each of the following sections we detail the probability of a given symptom in each of 5 phases of the disease.

\subsection{Fever, Chills}

\begin{table}[ht]
\begin{tabular}{lrr}
\hline
Disease/Phase                             & \multicolumn{1}{l}{Fever} & \multicolumn{1}{l}{Chills} \\ \hline
\cellcolor[HTML]{FFFFFF}\textbf{COVID-19} &                           &                            \\ \hline
\cellcolor[HTML]{EFEFEF}Incubation        & 0\%                       & 0\%                        \\
\cellcolor[HTML]{EFEFEF}Onset             & 20\%                      & 80\%                       \\
\cellcolor[HTML]{EFEFEF}Plateau           & 80\%                      & 50\%                       \\
\cellcolor[HTML]{EFEFEF}Post-plateau 1    & 0\%                       & 0\%                        \\
\cellcolor[HTML]{EFEFEF}Post-plateau 2    & 0\%                       & 0\%                        \\
                                          & \multicolumn{1}{l}{}      & \multicolumn{1}{l}{}       \\
\cellcolor[HTML]{FFFFFF}\textbf{Flu}      &                        &                        \\ \hline
\cellcolor[HTML]{EFEFEF}First day         & 70\%                      & 0\%                        \\
\cellcolor[HTML]{EFEFEF}Main              & 70\%                      & 0\%                        \\
\cellcolor[HTML]{EFEFEF}Last day          & 30\%                      & 0\%                        \\ \hline
\end{tabular}
\end{table}

\subsection{Gastroenteritis, Diarrhea, Nausea / Vomiting}

If $innoculum \textgreater 0.6$ then the probability of this person having Gastrologic symptoms on any day is $(innoculum-0.15)/4$, and if on any day they have Gastrologic symptoms, then for each following day they will also have this type of symptom.
\begin{table}[]
\begin{tabular}{llrr}
\hline
Disease/Phase                             & Gastro                                  & \multicolumn{1}{l}{Diarrhea} & \multicolumn{1}{l}{Nausea\_vomiting} \\ \hline
\cellcolor[HTML]{FFFFFF}\textbf{COVID-19} &                                         & \multicolumn{1}{l}{}         & \multicolumn{1}{l}{}                 \\ \hline
\cellcolor[HTML]{EFEFEF}Incubation        & \multicolumn{1}{r}{0\%}                 & 0\%                          & 0\%                                  \\
\cellcolor[HTML]{EFEFEF}Onset             & \multicolumn{1}{r}{Innoculum - 0.15}    & 90\%                         & 70\%                                 \\
\cellcolor[HTML]{EFEFEF}Plateau           & \multicolumn{1}{r}{(Innoculum-0.15)/4}  & 90\%                         & 70\%                                 \\
\cellcolor[HTML]{EFEFEF}Post-plateau 1    & \multicolumn{1}{r}{(Innoculum-0.15)/10} & 90\%                         & 70\%                                 \\
\cellcolor[HTML]{EFEFEF}Post-plateau 2    & \multicolumn{1}{r}{(Innoculum-0.15)/10} & 90\%                         & 70\%                                 \\
                                          &                                         & \multicolumn{1}{l}{}         & \multicolumn{1}{l}{}                 \\
\cellcolor[HTML]{FFFFFF}\textbf{Flu}      &                                         & \multicolumn{1}{l}{}         & \multicolumn{1}{l}{}                 \\ \hline
\cellcolor[HTML]{EFEFEF}First day         & \multicolumn{1}{r}{70\%}                & 50\%                         & 50\%                                 \\
\cellcolor[HTML]{EFEFEF}Main              & \multicolumn{1}{r}{70\%}                & 50\%                         & 50\%                                 \\
\cellcolor[HTML]{EFEFEF}Last day          & \multicolumn{1}{r}{20\%}                & 50\%                         & 25\%                                 \\ \hline

\end{tabular}
\end{table}

\subsection{Fatigue, Hard Time Waking Up, Unusual, Headache, Confused, Lost Consciousness}
Let A represent Age/200, C represent Carefulness/2, and let I represent Inoculum. Let F represent A+(I*0.6)-C. Let HW represent the symptom, Hard time waking up, and LC represent the symptom, Lost consciousness.

The probability of a person having fatigue as a symptom is given in the first column, while the probabilities of having the other symptoms given that this person has fatigue are given in the columns following the Fatigue column. In order to have unusual symptoms, the person must also be over 75 years of age; that is, the probabilities in the table given for unusual symptoms are the probabilities of having unusual symptoms given this person is over 75 years of age and is experiencing fatigue.

\begin{table}[]
\begin{tabular}{lrrllrl}
\hline
Disease/Phase                             & \multicolumn{1}{l}{Fatigue} & \multicolumn{1}{l}{HW} & Unusual & Headache & \multicolumn{1}{l}{Confused} & LC   \\ \hline
\cellcolor[HTML]{FFFFFF}\textbf{COVID-19} & \multicolumn{1}{l}{}        & \multicolumn{1}{l}{}   &         &          & \multicolumn{1}{l}{}         &      \\ \hline
\cellcolor[HTML]{EFEFEF}Incubation        & 0\%                         & 0\%                    & 0\%     & 0\%      & 0\%                          & 0\%  \\
\cellcolor[HTML]{EFEFEF}Onset             & F                           & 60\%                   & 20\%    & 50\%     & 10\%                         & 10\% \\
\cellcolor[HTML]{EFEFEF}Plateau           & F                           & 60\%                   & 50\%    & 50\%     & 10\%                         & 10\% \\
\cellcolor[HTML]{EFEFEF}Post-plateau 1    & 1.5F+(I-0.15)               & 60\%                   & 50\%    & 50\%     & 10\%                         & 10\% \\
\cellcolor[HTML]{EFEFEF}Post-plateau 2    & 2F+(I-0.15)                 & 60\%                   & 50\%    & 50\%     & 10\%                         & 10\% \\
                                          & \multicolumn{1}{l}{}        & \multicolumn{1}{l}{}   &         &          & \multicolumn{1}{l}{}         &      \\
\cellcolor[HTML]{FFFFFF}\textbf{Flu}      & \multicolumn{1}{l}{}        & \multicolumn{1}{l}{}   &         &          & \multicolumn{1}{l}{}         &      \\ \hline
\cellcolor[HTML]{EFEFEF}First day         & 40\%                        & 30\%                   &         &          &                              &      \\
\cellcolor[HTML]{EFEFEF}Main              & 80\%                        & 50\%                   &         &          &                              &      \\
\cellcolor[HTML]{EFEFEF}Last day          & 80\%                        & 40\%                   &         &          &                              &      \\ \hline
\end{tabular}
\end{table}

\subsection{Trouble Breathing, Sneezing, Cough, Runny Nose, Sore Throat, and Severe Chest Pain}
Let I represent Inoculum, and let C represent Carefulness/4. Let TB represent the symptom trouble breathing, ST represent the symptom sore throat, and SCP represent Severe chest pain.

The probability of having trouble breathing as a symptom is given in the TB column for the corresponding COVID-19 phases. For an individual to experience severe chest pain (SCP), the individual must also be extremely sick; that is, the probabilities given for having SCP as a symtpom are the probabilities of having SCP given an individual has trouble breathing and is extremely sick.

\begin{table}[]
\begin{tabular}{lrrllrl}
\hline
Disease/Phase                             & \multicolumn{1}{l}{TB} & \multicolumn{1}{l}{Sneezing} & Cough & Runny Nose & \multicolumn{1}{l}{ST} & SCP  \\ \hline
\cellcolor[HTML]{FFFFFF}\textbf{COVID-19} & \multicolumn{1}{l}{}   & \multicolumn{1}{l}{}         &       &            & \multicolumn{1}{l}{}   &      \\ \hline
\cellcolor[HTML]{EFEFEF}Incubation        & 0\%                    & 0\%                          & 0\%   & 0\%        & 0\%                    & 0\%  \\
\cellcolor[HTML]{EFEFEF}Onset             & (I/2)-C                & 20\%                         & 60\%  & 10\%       & 50\%                   & 40\% \\
\cellcolor[HTML]{EFEFEF}Plateau           & 2*(I-C)                & 30\%                         & 90\%  & 20\%       & 80\%                   & 50\% \\
\cellcolor[HTML]{EFEFEF}Post-plateau 1    & I-C                    & 30\%                         & 90\%  & 20\%       & 80\%                   & 15\% \\
\cellcolor[HTML]{EFEFEF}Post-plateau 2    & (I-C)/2                & 30\%                         & 90\%  & 20\%       & 80\%                   & 15\%
\end{tabular}
\end{table}

\subsection{Light/Moderate/Heavy Trouble Breathing}
The probability of having light/moderate/heavy trouble breathing is given by the probability of an individual having light/moderate/severe symptoms of COVID-19 and the probability of having trouble breathing as a symptom.

\subsection{Loss of Taste}
The probability of having loss of taste as a symptom of COVID-19 is 25\% during the onset phase, 35\% during the plateau phase, and 0\% for all other phases of COVID-19.

\subsection{Aches}
Aches are not caused by COVID-19 in this simulator, but are caused by the flu. The probabilities of having aches on the first and last days of the flu are 30\% and 80\% respectively, while the probability of having aches for all other days with the flu is 50\%.

\subsection{Informative Symptoms}
\label{app:informative-symptoms}

\section{Scaling Analysis} \label{sec:scaling}

\begin{figure}[htp!]
\centering
\begin{subfigure}{\textwidth}
  \centering
  \includegraphics[width=\linewidth]{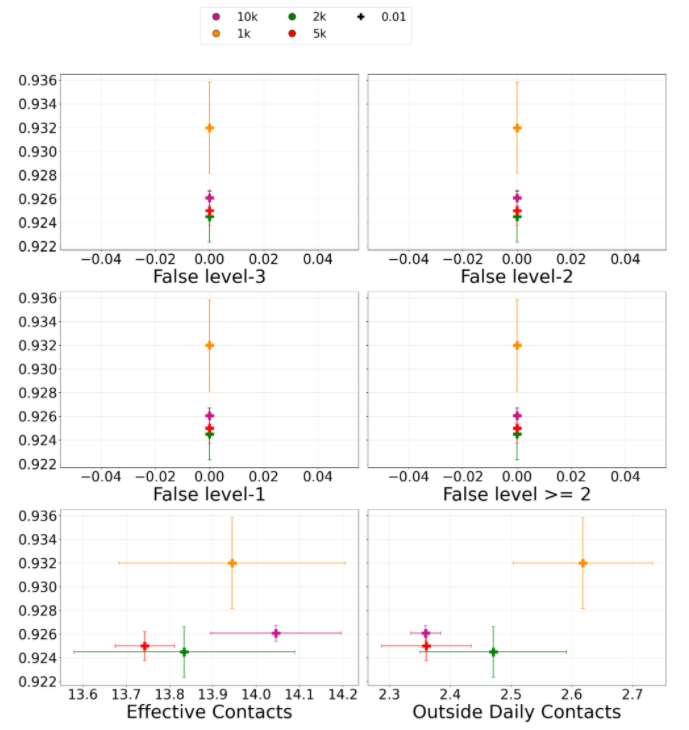}
\end{subfigure}
\caption{\textbf{Scaling analysis} verifying several metrics for different population sizes from 1k to 10k. We observe that for 2k and above, dynamics are qualitatively similar, while 1k is an outlier by several metrics. We thus use populations of 2k and above for all experiments.}
\label{fig:scaling}
\end{figure}

We wish to verify that the dynamics of the simulator at the population sizes we model are representative of larger populations. Because of the computational demands of an agent-based model, particularly with messaging between agents, it is more efficient to model smaller populations, as long as the dynamics remain reprsentative. As shown in Figure \ref{fig:scaling}, we find that population sizes above 2k to be representative of the dynamics of larger populations across a range of metrics. We thus ensure all experiments are run with populations of 2k and over.

\section{Cost-Effectiveness Calculations} \label{app:cost-effectiveness}
\subsection{DALYs}
Disability-Adjusted Life Years (DALYs) are a summary measure of the public health burden associated to a specific cause's premature mortality and morbidity. 
To calculate DALYs, we individually compute the years of life lost due to premature mortality (YLLs)\cite{YLLCalc} for agents that died during the simulation, as well as the years of life lost due to disability (YLDs)\cite{YLDCalc} for agents that were infected and symptomatic. Disability weights (DW) are taken from the 2017 Global Burden of Disease Study\cite{DW}: they represent health preferences such that a DW of 0 is perfect health and a DW of 1 is equivalent to death. Hence, the higher the amount of DALYs, the worse health outcomes are. DALYs are calculated without discounting or age-weighting, following the WHO methodology\cite{WHOnodiscount}.

\begin{table}[htp!]
\begin{tabular}{lllll}
                \hline \begin{tabular}[c]{@{}l@{}}Symptomatic \\ agent status\end{tabular} & \begin{tabular}[c]{@{}l@{}}Disability \\ weight\end{tabular} & Equivalent to                        &  &  \\ \hline
\cellcolor[HTML]{EFEFEF}not hospitalized & 0.051                                                        & moderate lower respiratory infection &  &  \\
\cellcolor[HTML]{EFEFEF}hospitalized     & 0.133                                                        & severe lower respiratory infection   &  &  \\
\cellcolor[HTML]{EFEFEF}Critical care    & 0.408                                                        & severe COPD without heart failure    &  & 
\end{tabular}

\caption{Disability weights drawn from the 2017 Global Burden of Disease Study. 
Since there are no published weights for COVID-19, weights for similar health conditions are selected. 
As can be seen, higher weights are associated with worse health states.}
\label{tab:DWtable}
\end{table}

\begin{align*}
\text{For each agent $i$,} \\
    YLL_i & = \mathbbm{1}_{\text{i died}} \times L_i\\
    YLD_i & = \sum_s^{S} DW_s \times l_{i,s} \\
    DALY_i & = YLL_i + YLD_i \\ \\
\text{where:} \\    \\
L_i  &:= \text{Life expectancy of agent $i$} \\
DW_s &:=  \text{Disability weight of state $s$} \\
l_{i,s} &:=  \text{duration of state $s$ for agent $i$} \\
    \mathbbm{1}_{\text{i died}} &:=
\begin{cases}
1 \qquad \text{  if agent $i$ died during the simulation} \\ 0 \qquad \text{  otherwise}
\end{cases}
\end{align*}

\subsection{TPL}

\textbf{TPL} Temporary Productivity Loss (TPL) is the loss in productivity due to absenteeism from work. To calculate TPL, we extract from the simulator the number of work hours that agents aged 25 to 65 years had to forego due to quarantine, taking care of a dependent (supervision), or illness to the point of being unable to work. We then multiply this quantity of foregone work hours by the 2019 average hourly wage in Montreal\cite{MTLwage} to obtain total TPL. We follow the methodology for calculating TPL presented in \cite{nurchis2020impact}.
\begin{align*}
\text{Aggregated across all agents,} \\
    n_{25:65} &=  n_{25:65}^{quarantine} + n_{25:65}^{supervision} + n_{25:65}^{illness} \\
    TPL &= n_{25:65} \times w_{median} \\ \\
\text{where: } \\ \\
n_{25:65}^a &:= \text{number of hours foregone due to $a$} \\
w_{average} &:= \text{average hourly wage for Montreal workforce}
\end{align*}

\begin{table}[]
\begin{tabular}{lclrlrl}
\hline
Tracing method                                                                          & \multicolumn{2}{c}{{\em No Tracing}}             & \multicolumn{2}{c}{{\em Test-based BCT}} & \multicolumn{2}{c}{{\em Heuristic-FCT}} \\ \hline
\textbf{Metric}                                                                         & \multicolumn{6}{c}{}                                                                            \\ \hline
\cellcolor[HTML]{EFEFEF}\begin{tabular}[c]{@{}l@{}}Foregone Work\\ (1000 Hours)\end{tabular} & \multicolumn{1}{r}{49.510}   & $\pm$0.416  & 57.556    & $\pm$0.765    & 76.620    & $\pm$1.263  \\
\cellcolor[HTML]{EFEFEF}\begin{tabular}[c]{@{}l@{}}TPL\\ (\$M)\end{tabular}              & \multicolumn{1}{r}{1.370} & $\pm$0.012 & 1.591  & $\pm$0.019  & 2.122  & $\pm$0.038
\end{tabular}
\caption{Foregone work hours and TPL for different contact tracing methods. TPL is obtained by multiplying foregone work hours by the average Montreal salary of 27.67. Foregone work hours are aggregated across the population for agents aged 25 to 65.}
\end{table}

\subsection{Assumptions and Limitations}

\paragraph{\textbf{Disability weights} Covid-19 is a novel disease, and there are no published disability weights for different levels of severity of the disease.
Therefore, we use similar conditions as proxies for different health states of the agents.
Agent hospitalization status is used as a proxy for actual health status, and can be divided into three categories: agents that are symptomatic and not hospitalized, agents that are hospitalized but not in critical are, and agents that are in critical care. 
The disability weights, as well as their equivalent causes in the GBD 2017 study can be found in Table \ref{tab:DWtable}}

\paragraph{\textbf{Full trajectory} Due to the computational strain of the message-passing within the simulations, observing the full trajectory under binary contact tracing and feature-based contact tracing is currently unfeasible.
Future work will consider longer simulations that reach a post-pandemic steady state.}

\paragraph{\textbf{PPL} In addition to Temporary Productivity Loss (TPL) due to absenteeism from work, cost-benefit analyses following the Human Capital Model (HCA) \cite{nurchis2020impact} typically include a Permanent Productivity Loss (PPL) due to premature mortality component. 
When longer simulations become possible, future work will take into account PPL. }

\paragraph{ \textbf{Delaying vs Preventing} To properly evaluate the impact of different tracing strategies on the socio-economic burden of a disease, it is important to evaluate whether the proposed strategies avert DALYs, or simply delay them~\cite{vaillancourt}. 
Such evaluations require a longer trajectory to arrive at a post-pandemic steady state, which we will consider in future work. }

\paragraph{ \textbf{Sensitivity Analysis} The focus of this paper is not the cost-benefit analysis of the outcomes of the simulator, but rather the simulator itself, a sensitivity analysis of the cost-benefit results has been relegated to future work.}

\subsection{Additional data}

\begin{table}[htp!]
\begin{tabular}{lrrrrrrrrr}
\hline
 & \multicolumn{3}{c}{{\em No Tracing}}                                     & \multicolumn{3}{c}{{\em Test-based BCT}}                                & \multicolumn{3}{c}{{\em Heuristic-FCT}}                                  \\ \hline
\textbf{Age}     & DALYs                                    & YLD                  & YLL                  & DALYs                & YLD                  & YLL                  & DALYs                & YLD                  & YLL                       \\ \hline
\cellcolor[HTML]{EFEFEF}0-9                              & 0.70                 & 0.70                 & 0.00                 & 0.64                 & 0.64                 & 0.00                 & 0.30                 & 0.30                 & 0.00                 \\
\cellcolor[HTML]{EFEFEF}10-19                            & 1.15                 & 1.15                 & 0.00                 & 1.13                 & 1.13                 & 0.00                 & 0.75                 & 0.75                 & 0.00                 \\
\cellcolor[HTML]{EFEFEF}20-29                            & 30.93                & 1.15                 & 29.79                & 26.85                & 1.14                 & 25.71                & 12.92                & 0.74                 & 12.18                \\
\cellcolor[HTML]{EFEFEF}30-39                            & 24.70                & 1.32                 & 23.37                & 19.93                & 1.29                 & 18.64                & 12.49                & 1.01                 & 11.48                \\
\cellcolor[HTML]{EFEFEF}40-49                            & 18.54                & 1.75                 & 16.79                & 20.88                & 1.71                 & 19.17                & 22.84                & 1.27                 & 21.57                \\
\cellcolor[HTML]{EFEFEF}50-59                            & 47.37                & 1.82                 & 45.55                & 50.79                & 1.78                 & 49.01                & 31.32                & 1.18                 & 30.14                \\
\cellcolor[HTML]{EFEFEF}60-69                            & 72.01                & 1.26                 & 70.75                & 71.38                & 1.16                 & 70.21                & 26.39                & 0.59                 & 25.79                \\
\cellcolor[HTML]{EFEFEF}70-79                            & 155.87               & 1.92                 & 153.94               & 106.63               & 1.89                 & 104.74               & 86.38                & 1.22                 & 85.17                \\
\cellcolor[HTML]{EFEFEF}80-89                            & 213.44               & 1.54                 & 211.90               & 203.94               & 1.49                 & 202.45               & 123.73               & 1.00                 & 122.73               \\
\cellcolor[HTML]{EFEFEF}90-99                            & 149.23               & 1.80                 & 147.42               & 122.42               & 1.70                 & 120.72               & 92.41                & 1.21                 & 91.20                \\
\cellcolor[HTML]{EFEFEF}100+                          & 87.27                & 2.16                 & 85.12                & 87.01                & 2.02                 & 84.99                & 67.36                & 1.46                 & 65.90                \\
                                                         & \multicolumn{1}{l}{} & \multicolumn{1}{l}{} & \multicolumn{1}{l}{} & \multicolumn{1}{l}{} & \multicolumn{1}{l}{} & \multicolumn{1}{l}{} & \multicolumn{1}{l}{} & \multicolumn{1}{l}{} & \multicolumn{1}{l}{} \\
\textbf{Sex}                                             & \multicolumn{1}{l}{} & \multicolumn{1}{l}{} & \multicolumn{1}{l}{} & \multicolumn{1}{l}{} & \multicolumn{1}{l}{} & \multicolumn{1}{l}{} & \multicolumn{1}{l}{} & \multicolumn{1}{l}{} & \multicolumn{1}{l}{} \\ \hline
\cellcolor[HTML]{EFEFEF}male                             & 26.36                & 1.39                 & 24.97                & 25.61                & 1.34                 & 24.27                & 13.05                & 0.89                 & 12.16                \\
\cellcolor[HTML]{EFEFEF}female                           & 44.53                & 1.39                 & 43.14                & 40.99                & 1.35                 & 39.64                & 30.01                & 0.91                 & 29.10               
\end{tabular}
\caption{YLL, YLD and DALYs $\times 1000$ for different CT methods. In all three cases, the bulk of DALYs is due to Years of Life Lost due to premature mortality (YLL), rather than Years of Life Disabled (YLD). Women are also disproportionately affected in all three scenarios.  }
\end{table}

\begin{figure}[htp!]
\centering

\includegraphics[width=1\columnwidth]{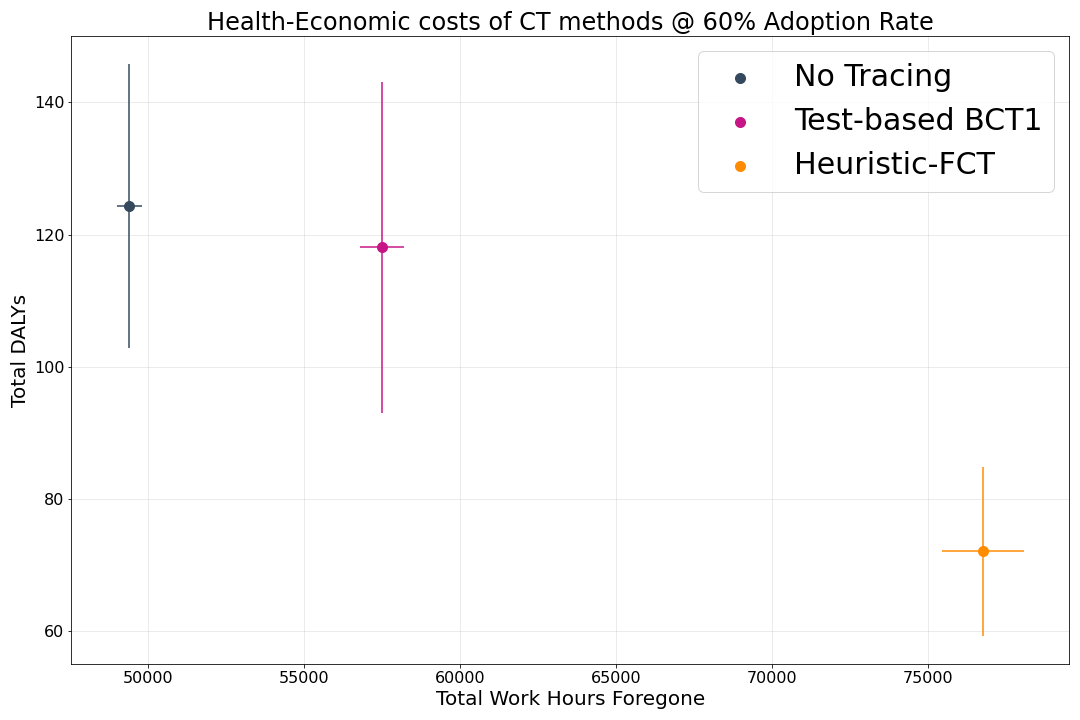}
\caption{
Impact of different CT methods on DALYs and total foregone work hours. 
{\em No Tracing} foregoes the least amount of work, but results in a high amount of DALYs. 
{\em Test-based BCT} foregoes more work, but still results in a large loss of health. 
{\em Heuristic-FCT} foregoes even more work than {\em No Tracing} does, but results in a large decrease of DALYs, alleviating the health burden. 
Total foregone hours due to quarantine, aggregated across individuals, are scaled by a factor of 0.49\cite{gallacher2020remote} to account for the proportion of agents able to work from home. 
Standard errors are computed by bootstrapping 100 samples of 6 runs over 10 seeds.
}
\label{fig:DALYsWork}
\end{figure}

As can be seen in Figure \ref{fig:DALYsWork}, under No Tracing, the total DALYs is 129.42, and the TPL is \$1.370M. {\em Test-based BCT} slightly affects the health and economic outcomes: the TPL is \$1.591M and the total DALYs is 119.42. However, {\em Heuristic-FCT} has a comparatively large effect on the total DALYs, which drops to 71.31. However, this is contrasted by a rise in the TPL to \$2.122M: health outcomes are drastically improved at the cost of a greater drop in productivity. Of note is the difference in impact of both tracing methods: whereas {\em Test-based BCT} has very little effect on the health and economic outcomes, {\em Heuristic-FCT} reduces total DALYs by 44.90\% at the expense of an increase in TPL by 54.89\%. 

\textbf{Bootstrapping} DALYs and TPL are computed for each seed separately before being aggregated to obtain a mean and standard error. When measures are calculated across subgroups of the simulation's population, such as across age groups, bootstrapping is used to capture \textbf{TODO}

\end{document}